\documentclass[graybox,natbib,nosecnum]{svmult}
\bibpunct{(}{)}{;}{a}{}{,} 

\pdfoutput=1   

\usepackage{mathptmx}       
\usepackage{helvet}         
\usepackage{courier}        
\usepackage{type1cm}        

\usepackage{xcolor}

\usepackage{makeidx}         
\usepackage{graphicx}        
\usepackage{multicol}        
\usepackage[bottom]{footmisc}
\usepackage[normalem]{ulem}	
\usepackage{hyperref}  
\usepackage{amsmath,amssymb, breqn}

\usepackage{soul}   
\def\ion#1#2{#1$\;${\small\rm\@{#2}}\relax}

\makeindex             

\begin{document}

\title*{Chemistry During the Gas-rich Stage of Planet Formation}
\author{Edwin A. Bergin and L. Ilsedore Cleeves}
\institute{Edwin A. Bergin \at University of Michigan,  Department of Astronomy, 1085 S. University Ave., Ann Arbor, MI, 48109, USA. \email{ebergin@umich.edu}
\and L. Ilsedore Cleeves \at Harvard-Smithsonian Center for Astrophysics, 60 Garden Street, 
Cambridge, MA 02138. \email{ilse.cleeves@cfa.harvard.edu}}
%
%
\maketitle
\abstract{
In this chapter we outline some of the basic understanding of the chemistry that accompanies planet formation.  We discuss the basic physical environment which dictates the dominant chemical kinetic pathways for molecule formation.   We focus on three zones from both observational and theoretical perspectives: (1) the planet forming midplane  and ice/vapor transition zones (snow-lines),
(2) the warm disk surface that is shielded from radiation, which can be readily accessed by todays observational facilities, and (3) the surface photodissociation layers where stellar radiation dominates.
We end with a discussion of how chemistry influences planet formation along with how to probe the link between formation and ultimate atmospheric composition for gas giants and terrestrial worlds.}



\bibliographystyle{spbasicHBexo}  

\section{Introduction}

Today with thousands of planets detected we are embarking on a new era of discovery.  Front and center will be the telescope time-consuming task of probing the chemical composition of exoplanetary atmospheres.     At the same time, the Atacama Large Millimeter Array (ALMA) has begun operation which has provided the first AU-scale resolved images of planet-forming disks.  This has led to a revolution in our understanding of the beginnings of planet formation.  
One of the fascinating areas to explore in the coming decade is how the final composition of a planet may be influenced by its formation environment.   The origin of water on our own planet is a key example, but similar questions exist regarding the disposition of water on Jupiter. To make these links, high signal-to-noise spectra of exoplanetary atmospheres and the continuing study of the solar system record at all scales are necessary.  However, we also need to have fundamental knowledge of how chemistry evolves before, during, and after planetary birth.  The focus of this chapter is to provide a baseline for the key chemical processes during the gas-rich stages of planet formation, including recent updates within this fast-moving field.

\begin{figure}[ht!]
\vspace*{-0.4 cm}
\centerline{\includegraphics[width=1.0\textwidth]{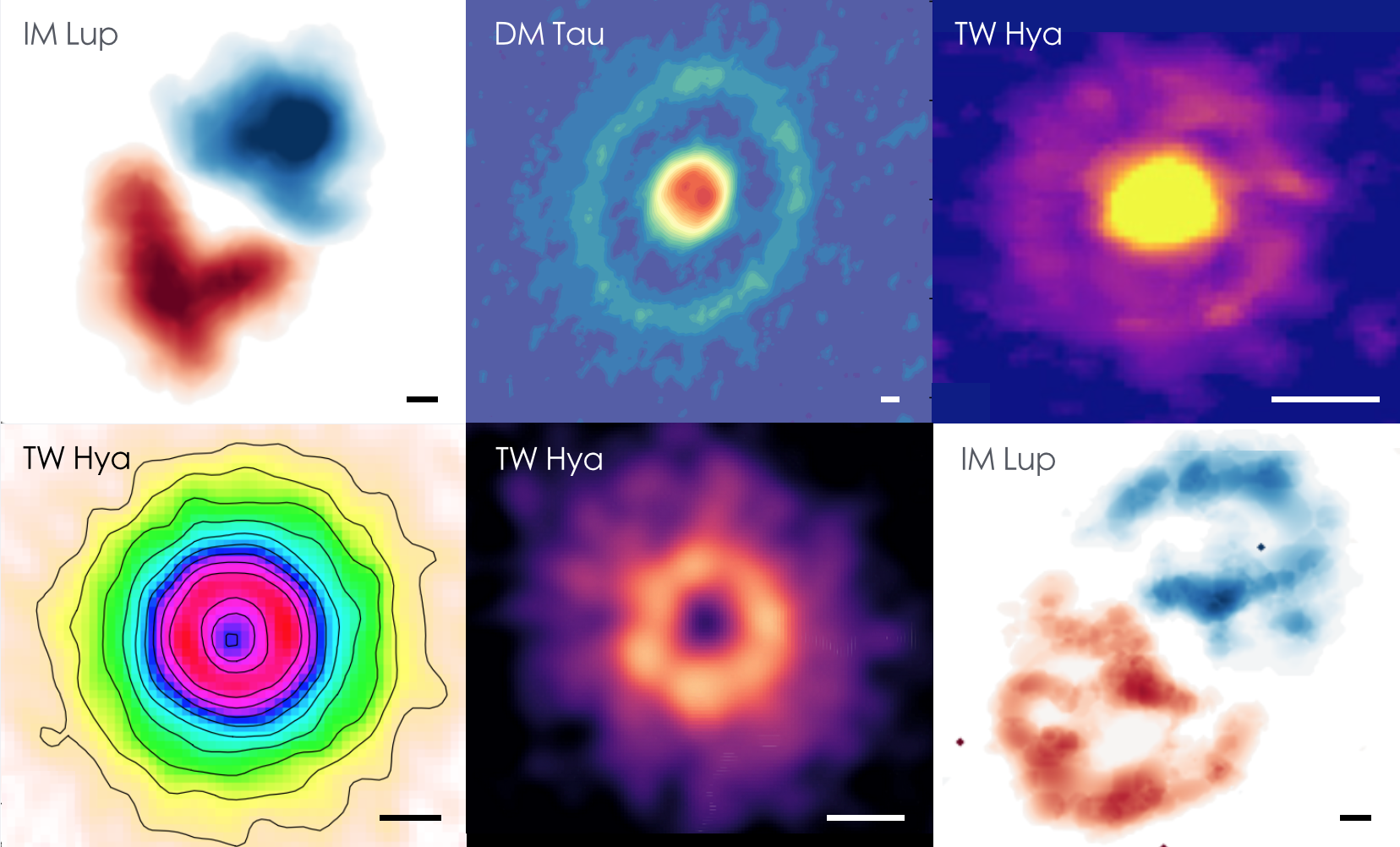}}
\caption{Molecular observations in disks exhibiting a wide variety of ring-like structure. Sources labeled in the top left and a 50 AU scale bar shown in the bottom right of each panel. From left to right and top to bottom: H$^{13}$CO$^+$ \citep{Oberg15a}, C$_2$H \citep{Bergin16}, $^{13}$CO \citep{Schwarz16}, CN \citep{Teague16}, N$_2$H$^+$ \citep[data from][]{Qi13_sci}, DCO$^+$ \citep{Oberg15a}, reproduced with permissions.}\vspace*{-0.2 cm}
\label{fig:montage}
\end{figure}

There have been a number of reviews of disk chemistry over the past decade \citep{bergin_ppv,henning13,Dutrey14}, and the goal of this chapter is not to provide a complete review of the field.  Rather, we aim our discussion towards new researchers in the relevant fields with references that link back to the grounding datasets, laboratory data, and/or theoretical underpinnings.   An important aspect of this volume is the discussion of new advances brought about by ALMA, which are still unfolding.  Fig.~\ref{fig:montage} presents a montage of molecular observations made with ALMA towards various disks of different masses, ages, and host stars.  One new aspect is the ubiquity of radial structure in the images with differences between species even when observed in the same source (e.g. C$_2$H, C$^{18}$O, CN, and N$_2$H$^+$ are each observed towards TW Hya).   This hints at an evolving and rich active chemistry which is responding to physical changes induced in the gas or the dust.

The chapter begins by detailing the evolving physical environment (dust/gas density and temperature, radiation field).  We then explore disk chemistry within this environment in three sections, in each case providing separate theoretical and observational perspectives.  These sections begin with the dense, shielded midplane with a strong focus on snow lines.  We then discuss the "warm molecular layer," where the conditions are warm enough such that molecules exist in gaseous form above a frozen midplane \citep{aikawa_vanz02}.  Finally, we explore the atomic-to-molecular transition at the disk surface itself.  We conclude by discussing how the disk composition might relate to overall planetary composition for both rocky and Jovian-like worlds.

\section{The Physical Environment}

\begin{figure}[ht]
\centerline{\includegraphics[width=.8\textwidth]{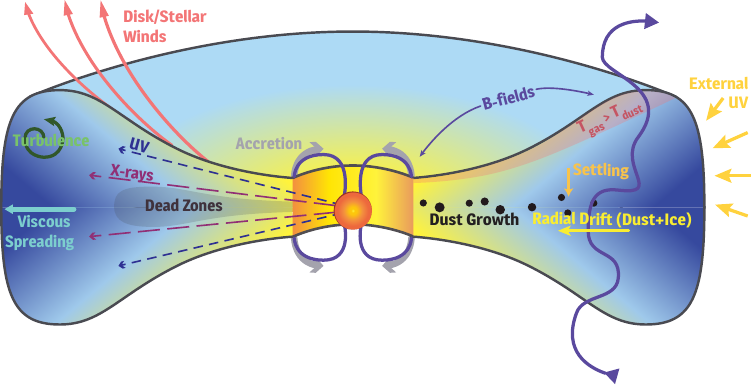}}\vspace{-2mm}
\caption{Illustration of key physical processes in protoplanetary disks. Reproduced with permission from Cleeves 2015 (PhD Thesis).}\vspace*{-0.2 cm}
\label{fig:phys}
\end{figure}

Decades of protoplanetary disk study have isolated several salient facts regarding the disk physical properties that have strong influence on the resulting chemical composition.  In Fig.~\ref{fig:phys} we show a generalized schematic of the relevant physical processes and their location(s) of influence.

\begin{figure}[ht]
\vspace{-4mm}
\centerline{\includegraphics[width=1.0\textwidth]{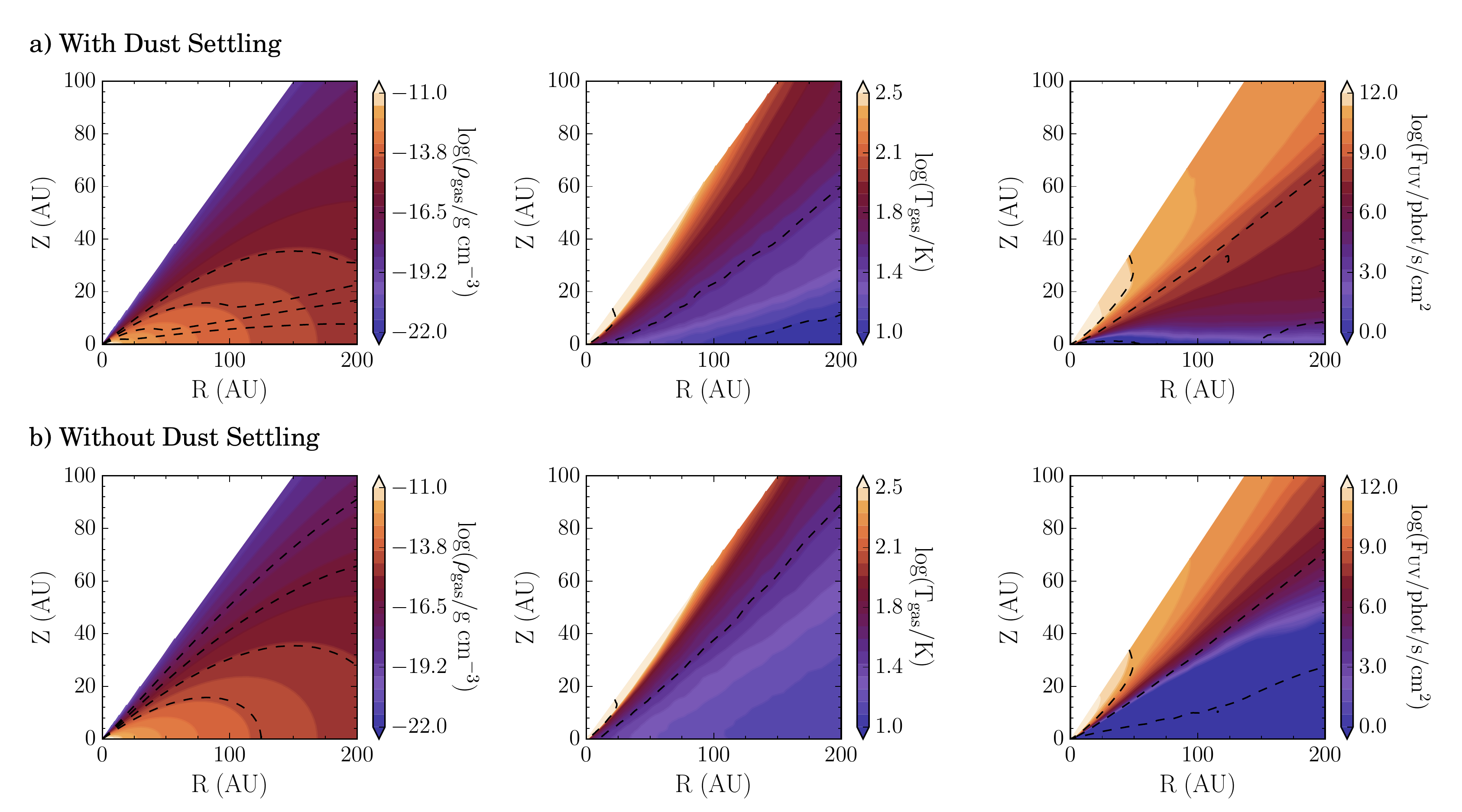}}
\caption{Example disk model for a gas-rich disk around a 1 M$_\odot$ star a) with settling (top) and b) without settling (bottom). Left panels are gas density, over-plotted with dust density contours as labeled. Middle panel is gas temperature overlaid with dust temperature contours. Right show the UV flux, overlaid with contours of the X-ray flux.}\label{fig:diskmod}
\end{figure}

\noindent {\em Gas Density Distribution:} early estimates of the gas density structure relied on a posterior distribution of planetary mass, adjusted by solar elemental abundances relative to hydrogen. This calculation naturally leads to a radial fall-off of the overall gas surface density profile as $\Sigma(R) \propto R^{-1.5}$, a distribution referred to as the ``minimum mass solar nebula'' \citep{hayashi_mmsn,w77_mmsn}.  To achieve consistency with the overall dust spectral energy distribution, modern models of disk evolution balance the effects of the irradiation of the central star dominated by the radiation at the blackbody peak, and that of hydrostatic equilibrium of the heated dust surface \citep{kh95,calvet91}.  A critical factor in this calculation is the disk viscosity, which is generally captured by the traditional $\alpha$ parameter which relates the turbulent viscosity ($\nu$) to disk motions, $\nu = \alpha c_s/\Omega_K$  \citep{ss73}.  Here $c_s$ is the isothermal sound speed and $\Omega_K$ the Keplerian rotation speed which will vary throughout the disk.  Fig.\ref{fig:diskmod} presents an example of a typical observationally motivated gas density structure. More detailed discussion regarding gas disk evolution and observational constraints can be found in the recent review of \citet{Bergin17}.

\noindent {\em Dust Density Distribution:} 
Observations of resolved sub-millimeter thermal dust emission continuum maps \citep[e.g.][]{Isella09, Andrews11}, the overall dust spectral energy distribution \citep[e.g.][]{furlan05}, and scattered light images \citep[e.g.][]{Debes13}  have now established that the dust has at least two major distributions.  Small (sub)-micron sized grains, responsible for much of the spectral emissions in the near and mid-infrared and for the absorption of high energy radiation, are dynamically coupled to the gas and have a spatial distribution that is comparable to that of carbon monoxide emission. As grains grow they are subject to differential forces that lead to  both settling and radial drift towards gas pressure maxima \citep[i.e. the midplane and the inner disk;][]{Whipple73, wc_ppiii,dd04}.  In the latter case, drift is prevalent for sizes that are large enough to be decoupled from the gas ($\ge$ 1 mm) but small enough that they feel effects of gas pressure ($<$ few km) depending on disk properties and orbital distance \citep{wc_ppiii}. Correspondingly, larger grains, which carry the disk dust mass have a distribution that is spatially concentrated. In the radial direction the large grains will drift into the inner tens of AU disk or perhaps greater distances, but are smaller than the CO/small grain scattered light disk.
In the vertical direction models suggest the grains exist in a very thin ($\sim$ 1 AU) layer \citep{Pinte16}.   
These issues are directly addressed in the chapter by Andrews \& Birnstiel. 

\smallskip

\noindent{\em High-Energy Radiation Field:} The main chemical pathways in dense gas at low temperatures ($\sim 10-100$~K) are through exothermic reactions between ions and neutrals \citep{hk73}.    Thus, the gas-phase chemistry is powered by ionizing sources. For disks there are  several important contributors to the overall ionization structure: UV and X-ray radiation (both internal and external) and galactic cosmic rays.  

\begin{enumerate}
\item {\em Galactic cosmic rays}, with an ionization rate in dense gas of $\zeta_{H2} \sim 5 \times 10^{-17}$ s$^{-1}$ \citep{Dalgarno06},  have the greatest penetration power, reaching gas at column densities up to $\Sigma_{CR} = 96$ g cm$^{-2}$ as given by \citet[][]{Umebayashi81} \citep[but see updates in][]{Padovani09,cleeves13a}. Recent work by \citet{cleeves13a} shows that the lower energy cosmic rays responsible for ionization might be impeded by stellar winds in a similar manner as in our own solar system \citep{webber_cosmicray}.  Winds can reduce the net low energy particle flux by many orders of magnitude, reducing the cosmic ray ionization rate by one or more orders of magnitude \citep{cleeves13a}.  

\item {\em X-rays} are an important source of deep ionization, while providing a steady productive source of ionization for the disk surface. Young low-mass stars, such as our Sun, are uniformly associated with stellar magnetic activity and active solar-like X-ray generating corona, and perhaps accretion.  The X-ray luminosity can vary with median values of $\sim L_X \sim 10^{30}$ s$^{-1}$ \citep{Guedel04} with typical X-ray temperatures of $kT_{X} \sim 1$ keV \citep{Preibisch05}.  Note there is also harder X-ray component with $kT_{X} > 3$~keV that may be associated with flare activity \citep{Preibisch05}.
X-ray photon propagation is governed by photoabsorption due to heavy elements \citep{mmc83,Henke93}, with scattering of increasing import above 5 keV \citep{ig99}.   For reference, normalized to solar abundances, the cross-section per hydrogen at 1 keV is $\sigma_{1\;keV} = 2 \times 10^{-22}$ cm$^{-2}$ while at 5 keV, $\sigma_{5\;keV} = 2 \times 10^{-24}$ cm$^{-2}$ \citep{mmc83, bb11b}.  As a consequence, there is a significant increase in penetration depth with energy \citep{ig99}.   Dust growth, settling, and drift can influence the X-ray radiation transfer with a minimum value set by volatile gas-phase absorbers \citep[e.g. H, C, O;][]{bb11b}.

\item {\em Ultraviolet Radiation} (UV) is dominated by accretion luminosity for low-mass cool stars (color temperatures $<$ 6000~K) and by the stellar photosphere for F stars and above.  The typical FUV luminosity for accreting (Classical T Tauri stars) young stars is $10^{-3} - 10^{-1}$~L$_\odot$; stars with little accretion, so called weak line T Tauri stars, have values of $L_{FUV} \sim  10^{-6} - 10^{-4}$ $L_\odot$ \citep{Yang12}.  For low mass stars the FUV luminosity has several components:  an underlying continuum attributed to the accretion shock \citep{gullbring00}, strong - but absorbed - photospheric Ly $\alpha$ emission \citep{herczeg_twhya1,Schindhelm12}, lines from highly ionized atoms \citep{Ardia13}, and a forest of H$_2$ emission lines \citep{France12}.  
Ultraviolet radiation has the smallest penetration depth $\sigma_{FUV} \sim 10^{-3}$~g~cm$^{-2}$, but it has a strong dependence on grain evolution as its propagation is governed by smaller $0.1$~$\mu$m grains that grow and settle to the midplane.
Thus, models of UV radiation transfer show that it is highly dependent on the photoabsorption and scattering properties of evolving dust grains \citep{vanz_etal01}.  Of particular importance is the strong Ly $\alpha$ radiation which dominates the FUV field containing 70-90\% of the FUV luminosity \citep{herczeg_twhya1,Schindhelm12}. Detailed models show that, due to H-atom scattering in the upper layers, Ly $\alpha$ photons have greater penetrating power than the other components of the FUV spectrum \citep{bb11a}; this has an imprint on the resulting chemistry \citep{fogel11}.

\end{enumerate}

\noindent {\em Temperature:} Stellar radiation and heating from accretion dominate the  thermal budget of disk systems. \citet{calvet91} demonstrated that this heating leads to a temperature inversion on the disk surface where the disk midplane is colder then the surface  with the superposition of an overall radial temperature gradient (e.g. Fig.~\ref{fig:phys} and Fig. \ref{fig:diskmod}).  This temperature profile has direct implications for the overall spectral emissions from dust \citep{cg97,dalessio1999,dalessio01} and also gas emission \citep{vanz_etal01, aikawa_vanz02}.   From the chemical perspective, the most general consequence is that there will be at least two sublimation fronts, e.g., snowlines, see the Midplane and Molecular Layer sections below. 
In addition, the evolution of the dust population (settling and drift) will also strongly effect the thermal structure, where a more settled disk tends to be globally warmer \citep[Fig. \ref{fig:diskmod} and see also][]{dcw05,cleeves2016,facchini2017}.

\section{Disk Chemistry by Environment}

Below we divide the disk into 3 vertical zones that delineate key chemical transitions in the disk.  These are as follows.  The disk {\bf midplane} is defined as the layer with the maximum of the gas density distribution in the vertical direction.  With high densities, fast gas-dust collision timescales, and decaying temperatures with distance from the star, this layer is dominated by sublimation fronts and gas/ice transitions in the radial direction.  The next vertical zone, the so-called "{\bf warm molecular layer}" \citep{aikawa_vanz02}, is defined by a chemical definition as the layer where the vertical dust temperature exceeds that required for sublimation of CO, the most volatile carrier of an abundant heavy element ($\sim$20~K). The top of this layer is defined by the zone where molecules will be photo-dissociated at $\tau_{uv} \sim 1$.  This layer has an active chemistry with ionizing agents, such as X-rays (and/or cosmic rays),  facilitating ion-molecule chemistry in UV-shielded gas.  The last vertical zone refers to the stellar-irradiation dominated surface and the "{\bf atomic-to-molecular transition}".  This layer is defined by having $\tau_{uv} \le 1$ whereupon there is an interplay between molecular dissociation and reformation of predominantly simple molecules with fast formation rates.

\subsection{Midplane}\label{sec:midplane}
\subsubsection{Theory}

The plane that intersects the geometrical center of the disk perpendicular to the rotation axis is referred to as the disk ``midplane''. The midplane is enhanced in dust that has grown and settled from the surface. In the midplane, the dust can further dynamically evolve and grow and eventually form planetesimals (see Andrews \& Birnstiel chapter). Thus the composition and conditions of the midplane are of particular interest for understanding the initial formation properties of planetesimals. 

The midplane is relatively cooler and more shielded from radiation compared to the disk surface at a given radius. The temperature in the midplane is regulated by gas accretion close to the star (within one to a few AU), and from reprocessed stellar radiation intercepted by small dust grains in the upper layers of the disk as discussed above. These small dust grains indirectly heat the disk midplane beyond a few AU. 
The densities are typically sufficiently high such that the gas temperature is equal to that of the dust due to frequent collisions between the two populations. 

Of particular importance to the midplane composition are the strong chemical transitions from the ice to vapor, or snowlines. Such locations have been posited as important sites for dust grown or 
planet formation \citep[e.g.,][]{Stevenson88}.
At sufficiently cold temperatures, the gas begins to ``freeze-out'' onto dust surfaces once the rate of molecules hitting the surface matches the rate of molecules thermally desorbing (sublimating) from the surface. Following the formalism of \citet{hkbm09}, the thermal desorption rate ($R_{td}$) per atom or molecule is given by,

\begin{equation}
R_{td,i} \simeq \nu _i e^{-E_{a,i}/kT_{gr}}.
\end{equation}

\noindent Here, 
$\nu_i= 
1.6\times 10^{11}\sqrt{(E_{a,i}/k)/(m_i/m_{\rm H})}$ s$^{-1}$ is the vibrational
frequency of the species in the surface potential well with $m_i$ and $m_h$ the mass of species $i$ and hydrogen; $k$ is the Boltzmann constant. $E_{a,i}$ is the binding energy of species $i$ to the surface.  To determine the sublimation temperature of a molecule from a given surface we balance the desorption rate with the flux of molecules that are absorbing from the gas ($F_{ab}$).

Thus,

\begin{equation}
F_{ab,i}\equiv N_{s} R_{td,i}f_{s,i} = 0.25 n_i v_i S,
\end{equation}

\noindent N$_s$ equal to number of surface sites available per cm$^2$ ($N_{s} \sim 
10^{15}$ sites cm$^{-2}$, for a 0.1 $\mu$m grain) and f$_{s,i}$ the fraction of those sites occupied by species $i$.   $n_i$ is the space density of species $i$, $v_i$ its thermal velocity, and S is the sticking coefficient which is generally assumed to be unity.  Solving for the grain temperature we will determine the sublimation temperature, $T_{sub}$ for a given species:

\begin{dmath}
T_{sub,i} \simeq \frac{E_{a,i}}{k}\left[ 57 + \ln \left[ \left({N_{s,i} \over {10^{15}\ {\rm 
cm^{-2}}}}\right)
\left({\nu _i \over {10^{13}\ {\rm s^{-1}}}}\right)\left({{1\ {\rm cm^{-
3}}}\over n_i}\right)
\left({{10^4\ {\rm cm\ s^{-1}}}\over v_i}\right)\right]\right]^{-1}.
\end{dmath}

Because desorption is balanced by the absorbing rate the sublimation temperature (for a given binding energy) will have a dependence on the local gas pressure.  In addition,
a particular molecule's adsorption energy can additionally depend on the properties of the grain surface. 
For example, highly polar water increases the binding energy of CO to water ice to a value of $E_a\sim1500$~K \citep{fayolle16}, compared to CO bound to pure CO ice, which has $E_a=855$~K \citep{Oberg09}. For a pressure of $10^{-10}$~bar, this corresponds to a difference in absolute desorption temperature of 19~K and 26~K, respectively. 

\begin{figure}[ht]
\vspace{-9mm}
\centerline{\includegraphics[width=1\textwidth]{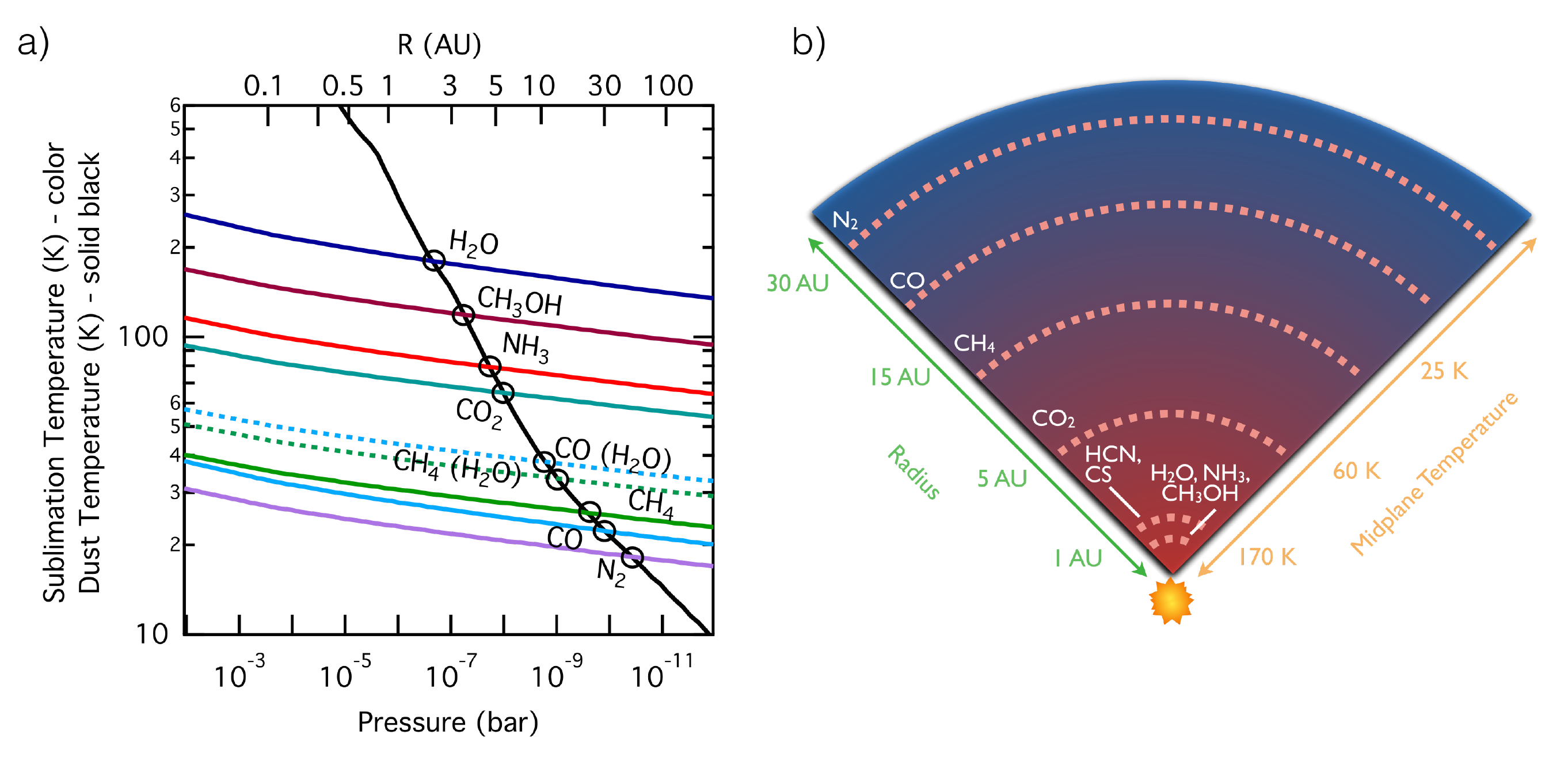}}
\caption{(Left:) Sublimation temperature of various molecules onto surfaces dominated by pure ices (i.e. same species) or water as a function of gas pressure.    Binding energies are from laboratory work with the following references,  N$_2$: \citet{fayolle16}, CO: \citet{fayolle16}, CH$_4$: \citet{Herrero10}, CH$_4$-H$_2$O: Behmard et al. 2018 (in prep.), CO-H$_2$O: \citet{Cleeves14a}, CO$_2$: \citet{Martin-Domenech14}, NH$_3$: \citet{Martin-Domenech14}, CH$_3$OH: \citet{Martin-Domenech14}, H$_2$O: \citet{fraser_h2obind}.
The solid line is one realization of a pressure-temperature profile in the midplane taken from \citet{dalessio06}. 
(Right:) Illustration of midplane freeze-out of various molecules with distance from the star (decreasing temperature).}\vspace*{-0.2 cm}
\label{fig:snowlines}
\end{figure}

Figure~\ref{fig:snowlines} illustrates the interdependence of gas pressure and dust temperature in setting location of various molecules' snowlines, along with a ``typical'' midplane temperature profile for a disk around a solar-mass star. Molecules such as H$_2$O, CH$_3$OH, and NH$_3$ have the highest $E_a$, and thus tend to remain in the ice phase until the temperatures are very warm. Molecules such as N$_2$ and CO have smaller $E_a$, and correspondingly they remain in the gas phase for tens of AU in radius for a solar mass star. 

\subsubsection{Observations}

\begin{figure}[ht]
\vspace*{-0.5 cm}
\centerline{\includegraphics[width=1\textwidth]{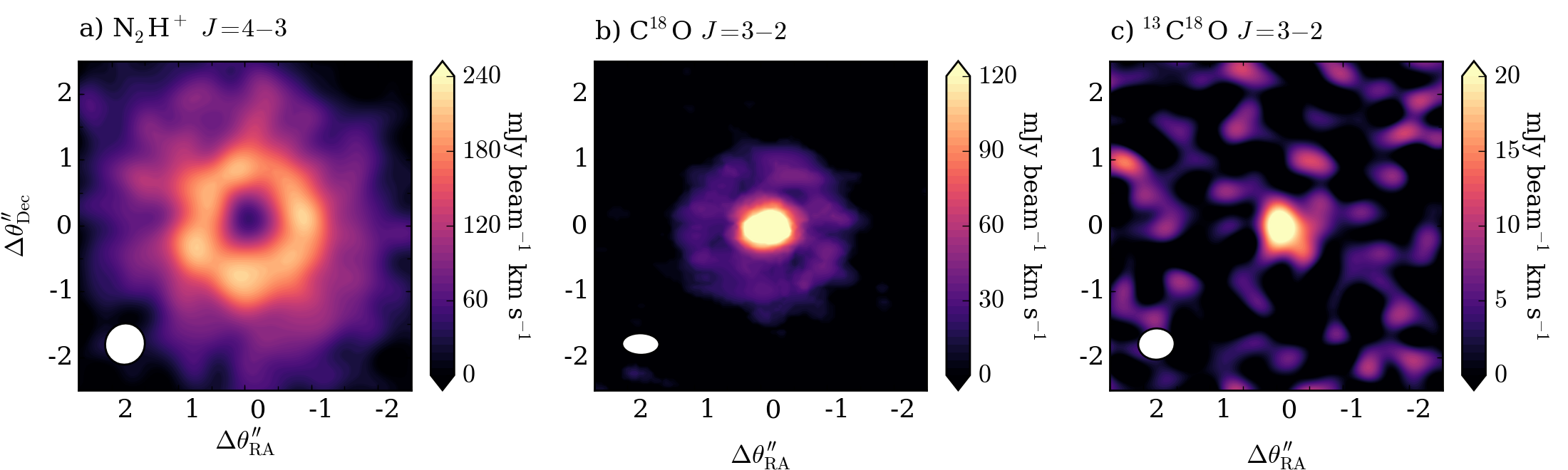}}
\caption{Observational tracers of the CO snow line. Data from \citet{Qi13_sci}, \citet{Schwarz16}, and \citet{Zhang17}, left to right.}\label{fig:snowdata}
\end{figure}

Most of the midplane gas is observationally ``hidden.'' The temperatures are too low in the outer disk (beyond tens of AU) for most observable species to remain in the gas, while those that do are often so cold the observable transitions are not strongly excited, or are obscured by warmer gas in the surface. In the inner disk, the high column densities present in disks make many observable transitions optically thick, also acting to ``hide'' midplane gas from our view. Furthermore, dust opacity from settled millimeter grains can act to obscure molecular emission from the midplane itself. One exception is that of $^{13}$C$^{18}$O whose low column density ($\sim30,000$ times less than $^{12}$CO, allows it to be observed down to the midplane inside of its snow line \citep{Zhang17}, see also Figure~\ref{fig:snowdata}.

There have been many major observational efforts to isolate the solid-to-gas phase transitions (snowlines) in the midplane. At present, the most observationally accessible snowline is that of CO. This feature is a direct consequence of CO's low $E_a$, and thus the CO snowline occupies tens of AU scales that are more readily observationally accessible than that of, e.g., H$_2$O (which will lie within $\sim$1~AU for a solar type star. The primary tracers of the CO snowline that have been used in the literature include N$_2$H$^+$, DCO$^+$, and optically thin isotopologues of CO. 

\noindent{\it N$_2$H$^+$:} The chemistry of N$_2$H$^+$ is such that it is rapidly destroyed in the presence of CO. As a consequence, N$_2$H$^+$ has been historically used as a marker of CO freeze-out in the dense interstellar medium \citep{charnley1997,Bergin02}. N$_2$H$^+$ shows a clear ring-like distribution in the TW Hya protoplanetary disk (see Figure~\ref{fig:snowdata}) with an inner radius of $\sim30$~AU.  Correspondingly, the \citet{Qi13_sci} paper interpreted the ring of N$_2$H$^+$ as a marker of the TW Hya disk's CO snow line, which was also in reasonable agreement with model temperature estimates of the disk at these radii \citep{Qi13_sci,qi2015}.


\noindent{\it CO Isotopologues:} More recently, measurements with ALMA of less abundant (and less optically thick) CO isotopologues found a steep drop off of C$^{18}$O intensity at $\sim17-23$~AU, interior to the N$_2$H$^+$ ring \citep{Schwarz16}. These results were more recently supported by $^{13}$C$^{18}$O observations, which place the column density break at a radius of $20.5\pm1.3$ AU \citep{Zhang17}. The break is attributed to freeze out at the midplane CO snow line. 

\begin{figure}
\vspace*{-0.5 cm}
\begin{center}
 \includegraphics[width=5in]{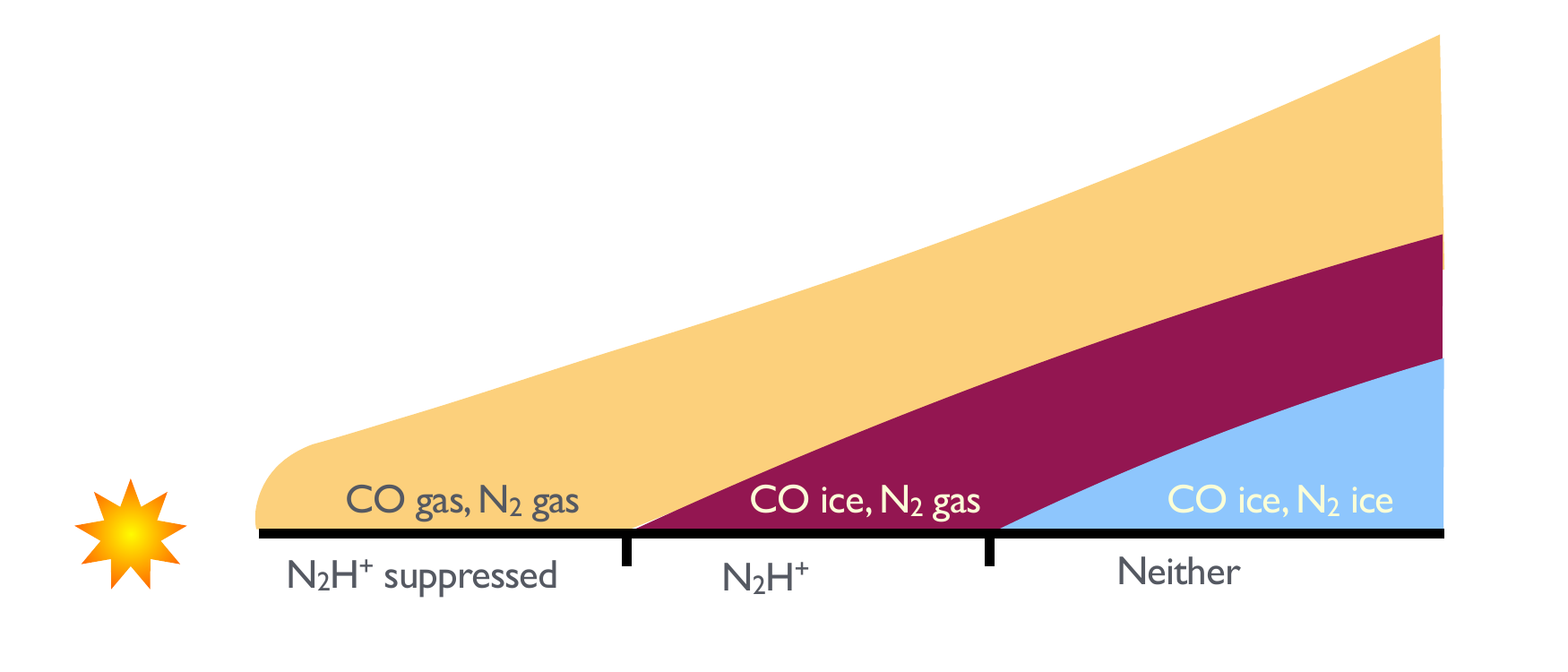} 
 \caption{Schematic of the ``classical'' picture of the relationship between N$_2$H$^+$ and the CO snow line. N$_2$ is expected to stay in the gas down to lower temperatures than CO, leading to a layer between the region of CO freeze out and N$_2$ freeze out where N$_2$H$^+$ is expected to occupy.}\vspace*{-0.8 cm}
   \label{fig:n2hdia}
\end{center}
\end{figure}

This $\sim10$~AU spatial discrepancy between the N$_2$H$^+$ transition and the CO transition is in part due to the inescapable nature of the disk temperature gradients, which are not purely radial. Instead, the vertical increase in temperature with height in the disk (see Figure~\ref{fig:n2hdia}) causes the region of CO freeze-out to occupy a wedge, with N$_2$H$^+$ in the simplest case bounding its borders until N$_2$ freeze-out commences. More detailed modeling of the chemistry \cite[e.g.,][]{aikawa2015,vanthoff2017} shows that the distribution of N$_2$H$^+$ is complicated by additional factors, including the desorption rates of CO vs. N$_2$, CO abundance, and disk ionization. 

\subsection{Molecular Layer}
\label{sec:wml}
\subsubsection{Theory}

In between the extremes of freeze-out and dissociation by ultraviolet photons, there exists a layer rich in gas-phase molecules, the ``warm molecular layer'' \citep{aikawa_vanz02}. This layer does not have discrete boundaries, but rather contains a mixture of neutral and ionized molecules and more strongly bound ices like water and ammonia. It is shielded sufficiently from strong UV irradiation to not fully destroy newly formed molecules, yet enough UV penetrates to facilitate a rich radical-driven chemistry. Within this layer, the disk transitions from being optically thin to X-rays to becoming thick, and can sustain a rich ion-neutral gas phase chemistry. Simultaneously, the temperatures are becoming cool enough such that molecules adhere to the surface of the cold dust grains ($T_{\rm dust} < 50$~K) long enough to initiate grain surface chemistry \citep[e.g.,][for H$_2$CO]{loomis2015}. In terms of building toward molecular complexity, these factors make the warm molecular layer a highly chemically active region in disks.

\begin{figure}[t]
\vskip2pt
\centerline{\includegraphics[width=1\textwidth]{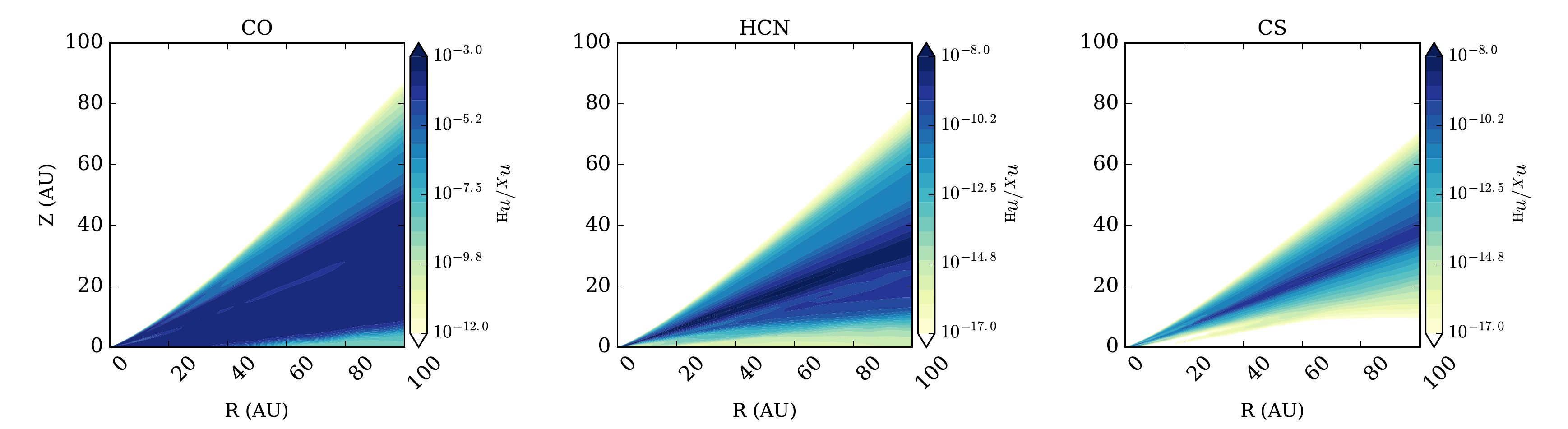}}
\caption{Example molecular distributions of key molecules known to be abundant in disks, CO, HCN, and CS. Model physical structure from Figure~\ref{fig:diskmod}.  Taken from \citet{cleeves2016}.}
\label{fig:molecules}
\end{figure}

Figure~\ref{fig:molecules} shows typical chemical model abundances for three commonly observed molecules in disks, CO, HCN, and CS, for a disk around a solar mass star \citep[taken from][]{cleeves2016}. CO is the second most abundant small molecule in the interstellar medium (after H$_2$), which is in part a testament to its robust chemistry. As is shown in the Figure, CO is present in the gas phase for a large radial and vertical portion of the disk. Over time, CO tends to form molecules like CH$_3$OH and CO$_2$, which have higher binding energies. These molecules freeze-out onto small dust grains removing some amount of CO \citep[see, e.g.][]{Bergin14, Furuya14, Reboussin15}. This behavior produces some of the banded structure in Figure~\ref{fig:molecules}. 

All three molecules have a steep abundance drop off in the upper atmosphere, traced by $z/r \gtrsim 0.5$, corresponding roughly to the $\tau_{\rm UV}\sim 1$ surface. At the lower bound, the molecules begin to freeze-out. In these particular models, the assumed binding energy of CS is greater than HCN, causing CS to freeze out at higher altitudes than HCN, placing it in a more narrow layer. These illustrative cases demonstrate that the vertical distribution of molecules is very sensitive to the temperature and density gradients, through freeze out and high energy radiation opacity.


\begin{figure}[b]
\vspace{-0.3cm}
\centerline{\includegraphics[width=0.7\textwidth]{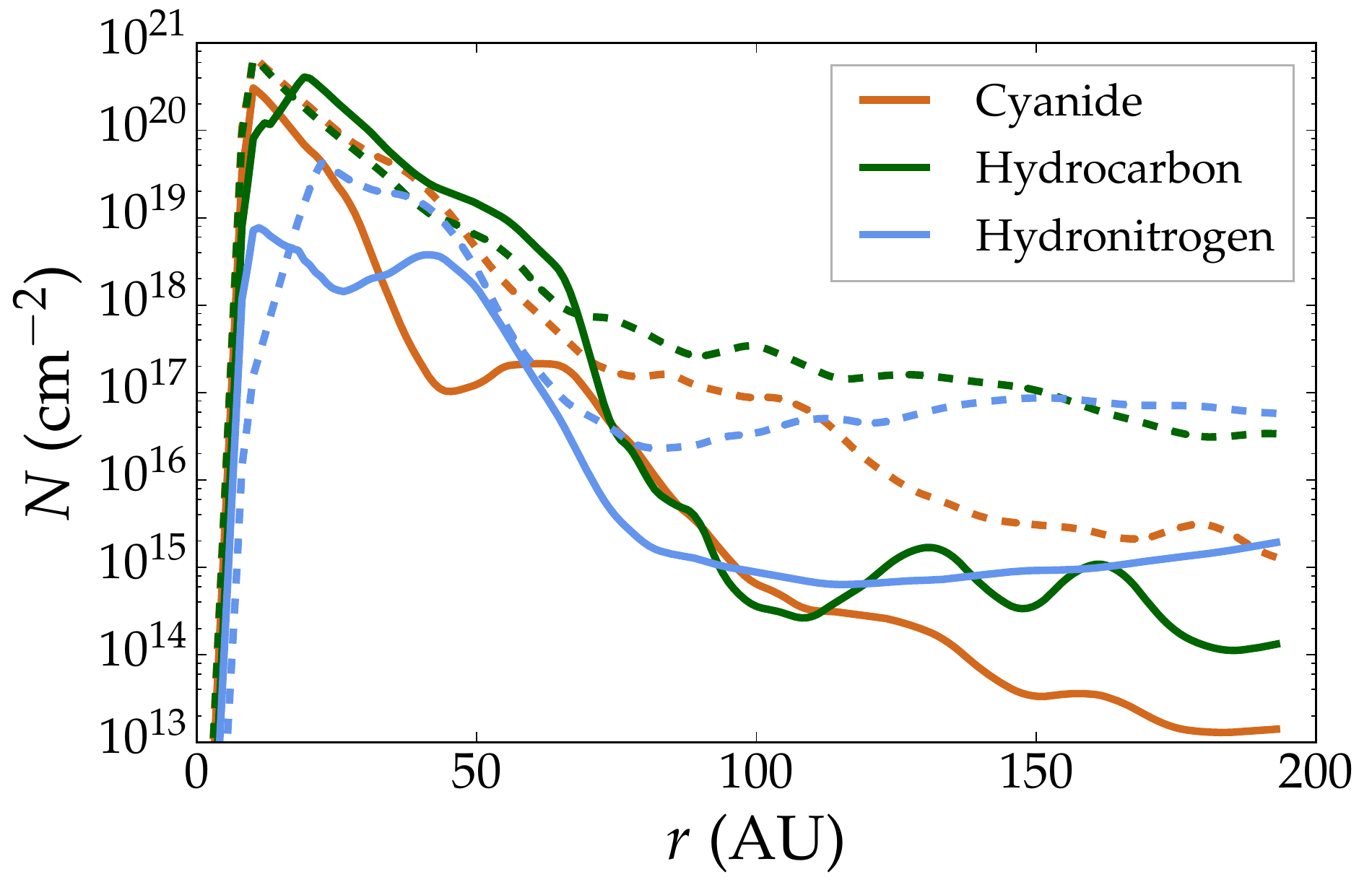}}
\caption{Simulated column densities with radial distance from the star for the molecular families (molecules with C-N bonds, C-H bonds, and N-H bonds) as indicated in the legend. The solid lines show models with interstellar carbon and oxygen abundances, while the dashed lines show models that have reduced carbon and oxygen abundances. Reproduced with permission from \citet{Du15}}\vspace{-0.3cm}
\label{fig:carbonmod}
\end{figure}

The molecular layer is perhaps most strongly subject to (and in some ways, responsible for) the time-evolving chemical effects of dust growth and settling. 
Ice-coating, especially that of water, aids in the probability of grains sticking together upon collision \citep[e.g.][]{wang2005}. As grains grow and decouple from the gas, the process of settling naturally displaces volatiles. In the absence of vigorous mixing, the process preferentially removes volatiles from the surface, sequestering the ices into the midplane, where they can further dynamically evolve. As water is a dominant ice constituent both in the interstellar medium, comets, and likely disks \citep{vandishoeckH2Oreview}, this process will tend to deplete oxygen to a greater extent than carbon, raising the carbon to oxygen ratio of the gas in the surface. As a result, the gas-phase chemistry will tend toward forming abundant carbon rich molecules (e.g., hydrocarbons and cyanides, see Figure~\ref{fig:carbonmod}).



\subsubsection{Observations}

\begin{figure}[b]
\centerline{\includegraphics[width=1.0\textwidth]{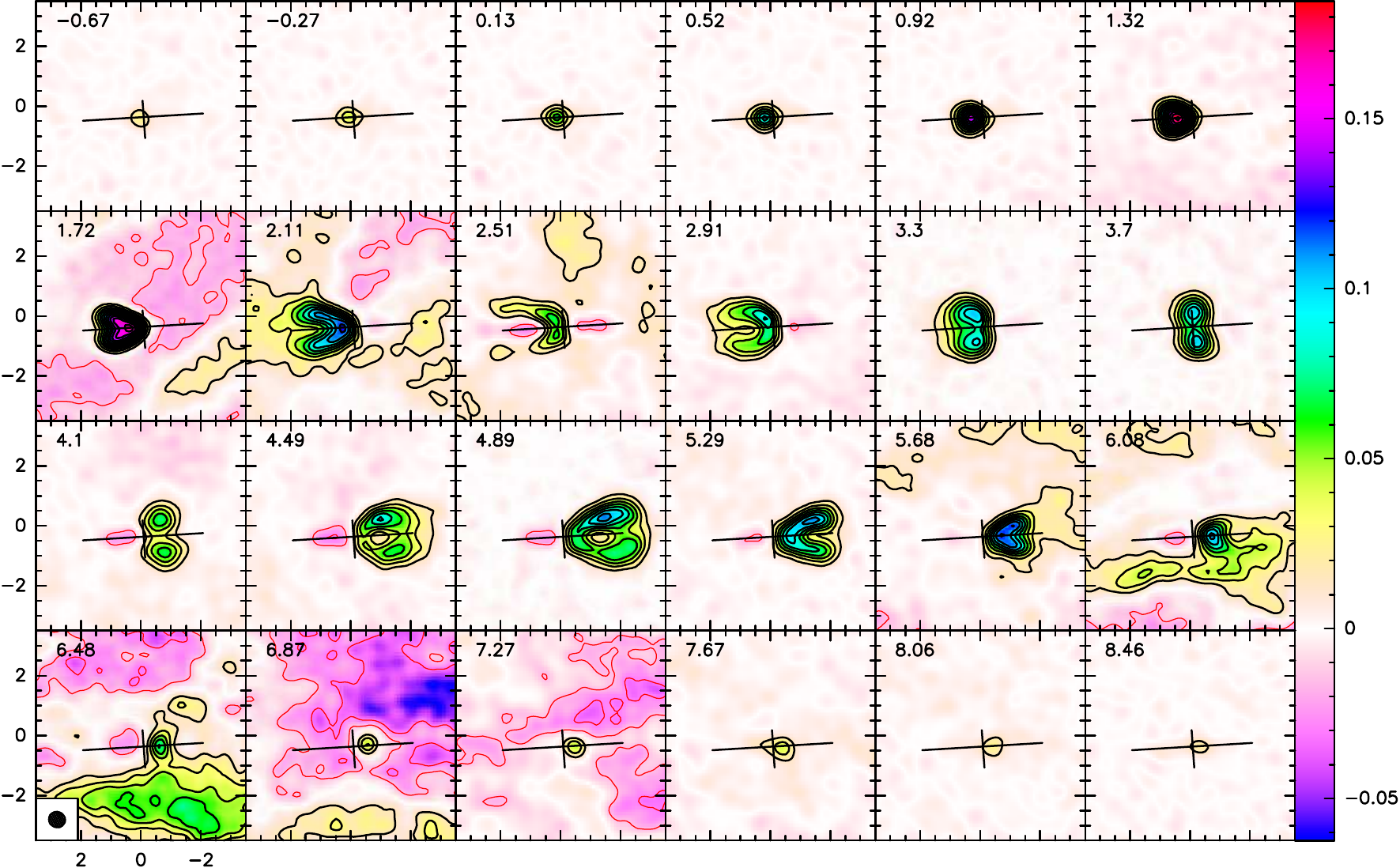}}
\caption{A protoplanetary disk viewed edge on in $^{12}$CO $J=2-1$. Reproduced with permission from \citet{dutrey2017}.  Each of these panels represents the emission seen in a velocity channel corresponding to the number in km s$^{-1}$ on the top right.  The systemic velocity is near 3.7 km s$^{-1}$.  Clear evidence is seen for rotation as the highest velocities correspond to the inner disk.  The horizontal line shows the midplane angle and note the absence of emission from distant disk layers ($>$ 1$''$) in the midplane at velocities near 2--3 km s$^{-1}$ and 4--5 km s$^{-1}$.  This missing emission is visual evidence for the CO-ice dominated zone.  }
\label{fig:flyingsaucer}
\end{figure}

As this region is rich in gas-phase molecules, most of our observations of the molecular content of disks come from the warm molecular layer. For edge-on disks, the vertical stratification can be directly observed (see Figure~\ref{fig:flyingsaucer}). 
In addition to CO, a host of additional molecules have been observed in disks. In the submillimeter, detected molecules include: CO, $^{13}$CO, C$^{18}$O, $^{13}$C$^{18}$O, HCO$^+$, DCO$^+$, H$^{13}$CO$^+$, HC$^{18}$O$^+$, CN, $^{13}$CN, C$^{15}$N, HCN, DCN, H$^{13}$CN, HC$^{15}$N,  CS, SO,  N$_2$H$^+$, N$_2$D$^+$, C$_2$H, C$_2$S, H$_2$CO, HC$_3$N,  $c-$C$_3$H$_2$, CH$_3$CN, CH$_3$OH. In the far-infrared, HD, H$_2$O, OH, NH$_3$, CH$^+$ and mid/near infrared, H$_2$, C$_2$H$_2$, H$_2$O, HCN, CO$_2$, and CH$_4$.
In addition, many of these molecules are not homogeneously distributed through the disk as demonstrated in  
 Figure~\ref{fig:montage}.

Consistent with the picture of time-depletion of oxygen relative to carbon, hydrocarbon rings have been demonstrated to be particularly bright \citep{Kastner15,Bergin16}. Thus, while the surface layers are undergoing active chemistry, along with volatile depletion, by studying the chemistry of the surface one can begin to learn indirectly about what is locked up, hidden from observations, in the midplane.


\subsection{Atomic-to-Molecular Transition}

\subsubsection{Theory}

The disk surface is highly exposed to energetic radiation from the star which is dominated by both the stellar far-UV (FUV: 912 \AA $\;< \lambda < 2000$~\AA) radiation and X-rays. This surface draws parallels with regions in the dense  ($>$ 10$^5$ cm$^{-3}$) interstellar medium that are irradiated by nearby massive stars, i.e., photodissociation regions or PDRs \citep{ht_rvmp}. Indeed, early work on chemical modeling of PDRs provide the fundamental physical basis (e.g., through the micro-physics of heating and cooling) for 
thermal-chemical models of disks 
\citep{Kamp04, gortihollenbach04, Nomura05,woitke09, Du14, Adamkovics14}. 

\begin{figure}[ht]
\vspace{-3mm}
\centerline{\includegraphics[width=1\textwidth]{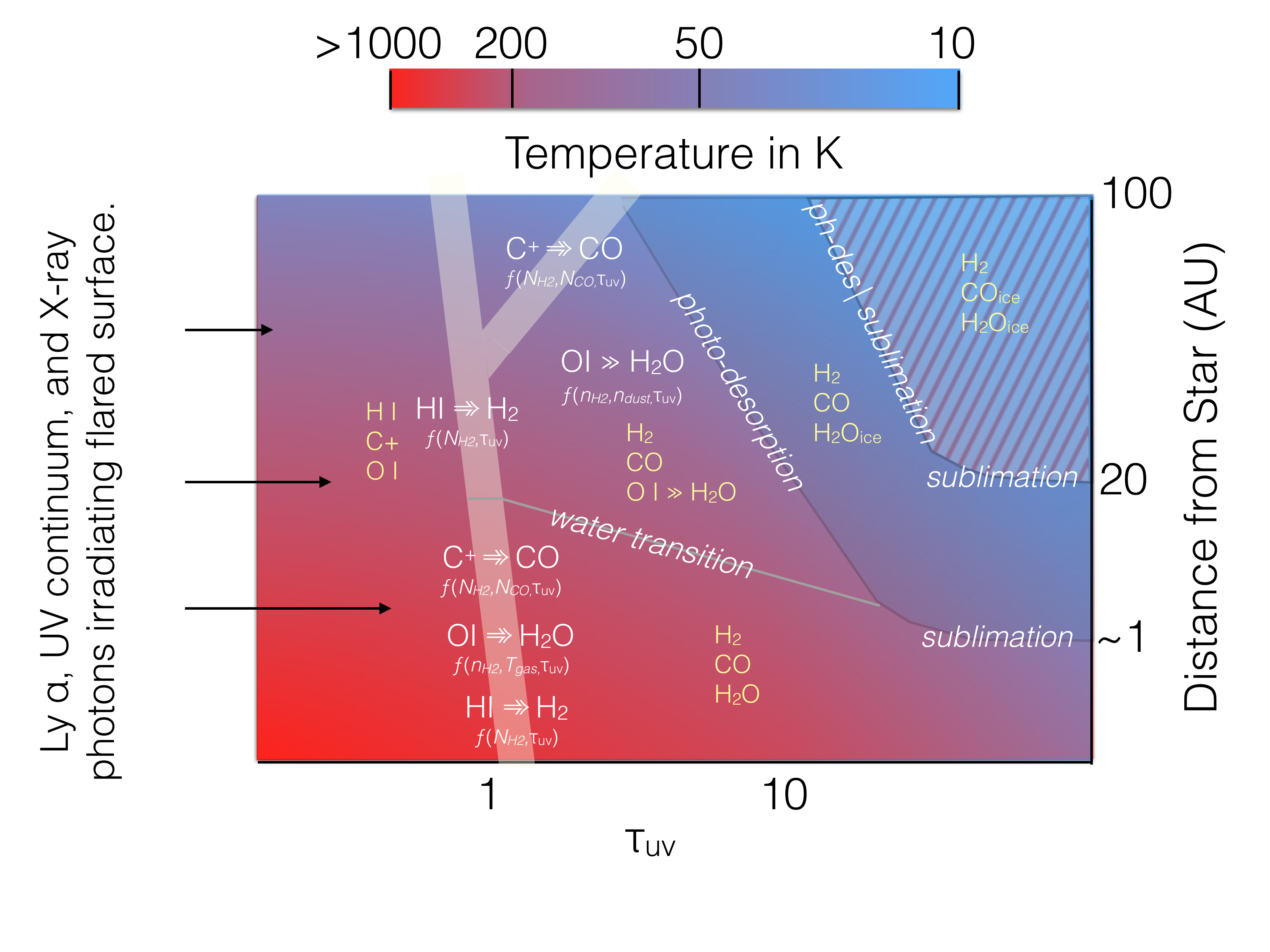}}\vspace{-3mm}
\caption{Schematic of the disk PDR surface illustrating key atomic-molecular transitions.  The width of the gray bars show the rough location of each transition.  H$_2$ and CO have transitions that are mediated by self-shielding. The water transition has two contributors.  In the inner disk its formation is dependent on the gas and dust temperatures due to fast neutral-neutral reactions in the gas and potential sublimation of icy grains.  Thus the O $\rightarrow$ H$_2$O transition does not solely depend UV optical depth.  Beyond a few AU and when the temperature falls below the water ice sublimation point, the fast gas-phase chemistry turns off and water is mostly as ice mediated by UV photodesorption.  In the outer ($> 100$ AU) disk photodesorption can also be important in the CO gas-ice transition when the temperature falls below its sublimation point and UV photons are present. The radial distances on the right ordinate are approximate and depend on the dust density profile and properties, the accretion rate, and the stellar spectral type. }\vspace{-3mm}
\label{fig:pdrT}
\end{figure}

In interstellar PDRs 
the strength of the radiation field is generally given in units of the local interstellar radiation field.  This is determined by computing the radiation generated by massive stars in the solar vicinity;  this has a value estimated to be $G_0$ = 1.6 $\times 10^{-3}$ erg~cm$^{-2}$~s$^{-1}$ \citep{habing68}. A typical UV radiation field of an accreting T Tauri star has a value of $G \sim 500$ $G_0$ at 100 AU \citep{bergin_h2,Yang12}. 
The dense (n $> 10^6$~cm$^{-3}$) disk close to the star thus has warm, T$_{gas} \sim 400 - 1000$~K, material in surface layer and a very hot disk photosphere ($>$ 1000~K) with cool ($\le$ 20~K) very dense material over a hundred AU from the star. This temperature and opacity profile produces important differences in the relative chemical transitions, compared to ISM PDRs, which are illustrated in Fig.~\ref{fig:pdrT} and summarized below.
  
\medskip
\noindent {\em \ion{H}{I} $\rightarrow$ H$_2$:} This transition is mediated by the impact of Ly $\alpha$ photons, which dominate the FUV radiation field and the reformation of H$_2$ onto warm sub-micron sized dust grains in surface layers.   A critical facet to understand for this layer, and for molecule formation in this transition zone in general, is the grain-surface formation of H$_2$ at high dust temperatures \citep{Cazaux02,ct04,Cazaux05}.      

\medskip
\noindent {\em \ion{C}{II} $\rightarrow$ C $\rightarrow$ CO:} Due to the high density of the inner disk  (n $\gg$ 10$^{7}$ cm$^{-3}$) this allows for rapid build-up of a CO self-shielding column.  Thus the \ion{C}{II} $\rightarrow$ C $\rightarrow$ CO transition can be commensurate with that of H$_2$ \citep{woitke09,Najita11}.  At larger distances from the star (many tens of AU), due to lower gas densities the carbon transition will shift away from H$_2$ and become more similar to an interstellar PDR.   In general, models use parametric descriptions of the CO photodissociation rate that incorporate shielding from CO, H$_2$, and dust \citep{visser09}.

\medskip
\noindent{\em \ion{O}{I} $\rightarrow$ OH $\rightarrow$ H$_2$O:} Depending the strength of the UV field and gas heating, in the inner disk this transition occurs when H$_2$ is present.   In surface layers the dominant formation mechanism for water vapor is via fast neutral-neutral reactions that rapidly process all free \ion{O}{I} into H$_2$O when T$_{gas} > 400$~K \citep{wg87,kn96}.  These reactions can become so fast that they can become competitive with photodestruction and H$_2$O can self-shield \citep{bb09}.   Water has photoabsorption cross-sections that are continuous in the FUV \citep{yoshino_h2oxs}; thus water can potentially shield other molecules from the destructive effects of UV in inner few AU \citep{bb09,Adamkovics14}.  Once the {\em gas} temperature falls below 400 K then the neutral-neutral gas phase pathways turn off.  Formation then follows the less efficient ion-molecule pathways linked to H$_3$O$^+$ and, in layers where T$_{dust} <$ T$_{sub}$(H$_2$O), requires UV photodesorption to release water and OH into the gas \citep{dchk05,hkbm09}.  The detailed review of the chemistry of water is presented in \citet{vandishoeckH2Oreview}. Thus there will be a radial transition on the disk surface from hot to cold chemistry where the water abundance drops precipitously by orders of magnitude.  The  surface transition is nominally distinct from the midplane snowline.  
 
 \medskip
\noindent {\em N, S, and metal ions:}  It is expected that the N to N$_2$ transition will be similar to that of CO, at least in behavior, as N$_2$ also can self-shield \citep{Heays14}.    The case of sulphur is more ambiguous as we have yet to locate the main repository of S in the dense interstellar medium and sulphur  appears to mostly be missing from the gas \citep{Druard12, Anderson13}.   A similar statement can be made about heavy metal ions (e.g., \ion{Fe}{II}, \ion{Si}{II}, \ion{Mg}{II}).   In the dense ISM these heavy elements appear to be locked in the refractory solid state  \citep{scs94, mb07} and the expectation is that this would also be the case in the protoplanetary disk.  If these species were present in abundance  they would be useful in mediating the interaction with the magnetic field \citep{Perez-Becker11,ilgner_diskmetals}.

\begin{figure}[ht]
\vskip2pt
\centerline{\includegraphics[width=1\textwidth]{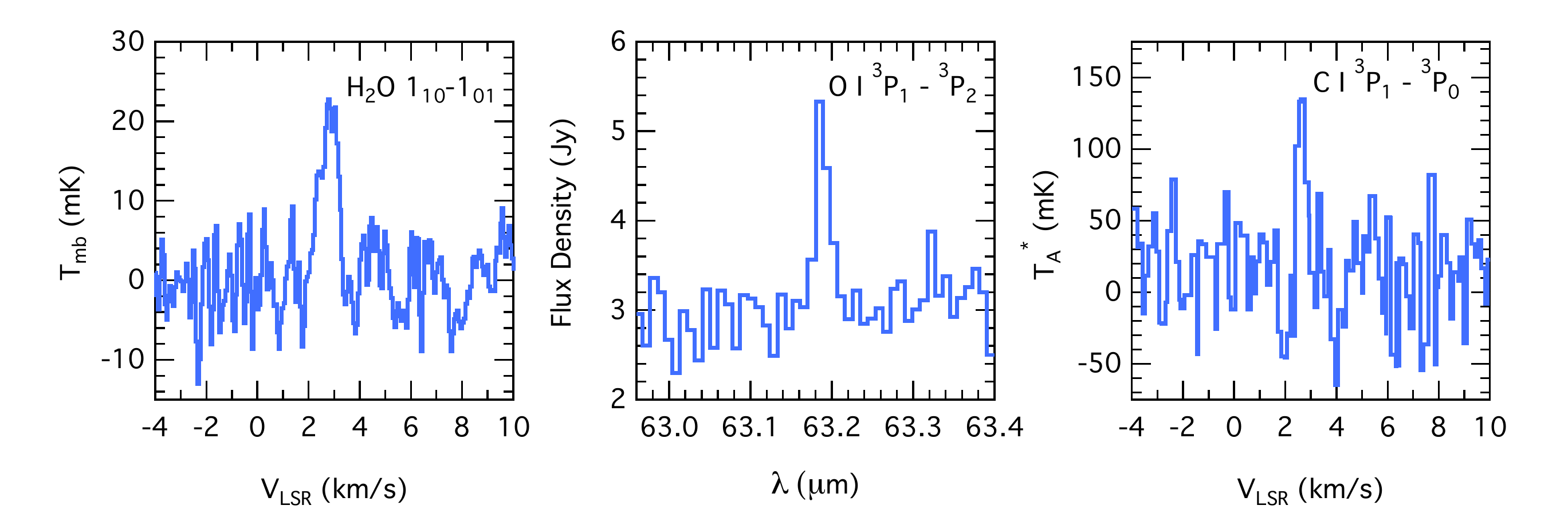}}
\caption{Sample observations towards TW Hya of \ion{O}{II} \citep{thi10}, \ion{C}{II} \citep{Kama16a}, and H$_2$O \citep{hoger11a}  ground state lines. Spectra obtained from Herschel Science Archive.}
\label{fig:pdrO}
\end{figure}

\subsubsection{Observations}
 
For the H$_2$ transition space telescopes such as {\it FUSE} and the {\it Hubble Space Telescope} can observe emission from the Lyman-Werner bands \citep{herczeg_twhya1, France12}.   Some of these emission lines coincide with the stellar Ly $\alpha$ emission and bear information on the true  Ly $\alpha$ spectral profile impinging on the disk surface \citep{herczeg_twhya1, France14}, which is important for models of chemistry \citep[e.g., ][]{fogel11} and  thermal balance \citep{Adamkovics14}.

Both \ion{C}{II} and \ion{C}{I} have emission lines that are observed via space-based platforms such as Herschel or via ground based observatories (for \ion{C}{I}).  In general, the detection statistics for  \ion{C}{II} are quite poor \citep{Howard13}, while \ion{C}{I} (Fig.~\ref{fig:pdrO}) has only been searched for in a handful of systems with moderate success \citep{Tsukagoshi15, Kama16a}. Low-J CO lines are readily observed, but generically trace deeper layers within the disk surface.

For oxygen, \ion{O}{I} has been surveyed with Herschel with a high rate of detection  \citep[e.g.][and references therein]{Howard13}. Analysis of the spectrally unresolved emission line shows that it is not clear whether the atomic oxygen emission traces the disk PDR or, in some sources, jets/outflows \citep{Alonso-Martinez17}.
OH and H$_2$O have rotational transitions in the mid-infrared and vibrational modes in the near-IR that probe warm ($\sim$few hundred K) surface layers.  Rotational line emission have been surveyed using {\it Spitzer} \citep{Salyk11,Pontoppidan11} and vibrational lines via ground-based observatories \citep{salyk08}.  Model analyses suggest that this emission is probing the PDR interior to the surface transition (Fig.~\ref{fig:pdrT} with a relatively high water abundance.  Lower energy rotational lines emitting in the far-infrared probe gas at larger radii where material has temperatures well below the sublimation temperature of water ice.  Thus, in this gas photodesorption is required to release water from grains\citep{hoger11a}.  
A deep survey using Herschel finds that detection rates are fairly low (below 10\%) with only two systems (TW Hya shown in Fig.\ref{fig:pdrO} and HD100546) having detectable emission lines \citep{du2017}.  This weak emission is interpreted as the result of a reduced water abundance in the photodesorbed gas \citep{hoger11a,du2017}, but see also \citet{Kamp13} for caveats regarding disk modeling uncertainties. Going forward, a combination of mid- and far-IR rotational emission can be used to constrain the surface water hot-cold transition \citep{zhang13,Blevins16}.

\section{Disk Composition and Planet Formation}

\subsection{Chemistry as a potentially controlling influence on planet formation}
The chemistry of the disk does far more than alter the composition of forming protoplanets, it can impact the formation process itself. As discussed in both the midplane and molecular layer theory, ice can impact the growth of grains by altering their sticking efficiency \citep[e.g.,][]{wang2005}. Correspondingly, the presence of water ice may enhance the rate of dust growth, seeding the initial stages of planet formation. It should be noted that not all ice enhances sticking. Experiments of CO$_2$ ice, for example, showed that CO$_2$ and silicate grains had similar collisional behavior in terms of growth and fragmentation \citep{musiolik2016,musiolik2016b}. Thus a fundamental understanding of the disk ice chemistry will be necessary for a complete treatment of dust evolution beyond 1 AU in disks \citep[e.g.,][]{Krijt16,stammler2017}.

The physical properties of the gas, such as viscosity, temperature, turbulence, and ionization fraction, too can impact the growth rate of solids and the later stages of planet formation. The accretion rate of gas onto planets, along with how planets sculpt the disk itself, is related to the disk viscosity. This very viscosity can additional act to mix the disk gas, which can alter its chemical composition by bringing together material from vastly different regions in the disk. As discussed above, the viscosity is often parameterized with $\alpha$. Typical accretion rates onto the star require high values of $\alpha\sim0.01$ \citep{hartmann1998}. For low mass disks, the classical expectation is that magnetic coupling with the weakly ionized disk generates magnetically driven turbulence through the magneto-rotational instability \citep[MRI;][]{balbus91}, which provides a source of disk viscosity.  

There have been substantial efforts to measure the predicted non-Keplerian, non-thermal motions in disks, such as those expected to be introduced by MRI \citep{simon2015} using observations of the bulk molecular gas, both unresolved \citep{hughes2011} and resolved \citep{Flaherty15,Teague16}. So far, highly constraining limits have been placed \citep[e.g., $<3\%$ of the local sound speed in the HD163296 disk;][]{Flaherty15}. Such estimates fundamentally depend on temperature, however, and thus caution must be taken when correcting out thermal motions versus turbulent motions \citep{Teague16}.


Nonetheless, such limits are an order of magnitude lower than MRI theory would predict, suggesting it is less efficient as a source of viscosity (and mixing) beyond 30 AU than previously thought. One possible explanation is that the disk may be more weakly ionized than expected, such that the gas and the magnetic fields are poorly coupled. \citet{Cleeves15} used observations of molecular ions in the warm molecular layer to infer the distribution of ionizing sources, and the corresponding net ionization fraction of the TW Hya disk. The distribution of molecular ions was found to be consistent with a low cosmic ray flux, and correspondingly, a low disk ionization fraction such that MRI should be inefficient for most of the disk midplane, out to $\sim50-60$~AU.



This tension between $\alpha$ and accretion rates onto the star have another potential solution, namely to have the angular momentum carried away instead by magneto-hydrodynamic disk winds \citep[e.g.][]{konigl2000,salmeron2007, suzuki2009, bethune2017, bai2016}, built off of the earlier seminal work of \citet{blandford1982}. However, these winds must not generate turbulence, especially in the upper layers where the observational constraints are most stringent \citep{simon2017}. Using a series of shearing box simulations, \citet{simon2017} investigated which parts of parameter space satisfy both high accretion and maintain a laminar disk state, and found that a net vertical magnetic field threading the disk is crucial to matching both conditions. Going forward, observational constraints on the disk magnetic field through, e.g., the Zeeman effect will help test this actively evolving theoretical picture.

\subsection{Global disk composition and its connection with gas giant planetary composition} 

In general, the bulk of planetary gas giant atmospheres are in chemical equilibrium \citep[except in specific instances, e.g.,photochemistry and transport induced quenching][]{Moses14}.  Thus the particular molecular form of delivery for say carbon (e.g. CH$_4$, CO, CO$_2$, hydrocarbons) will not matter; for a given pressure-temperature the expectation is that all carbon will reside in CH$_4$ for cool ($<$ 1000 K) atmospheres or CO for warmer conditions.   Thus, if we seek to explore connections between giant planet atmospheric composition with the composition of the {\em gas} at formation we must look at elemental pools and target the main gas phase carriers of abundant elements.   

\begin{figure}[h]
\begin{center}
 \includegraphics[width=4.3in]{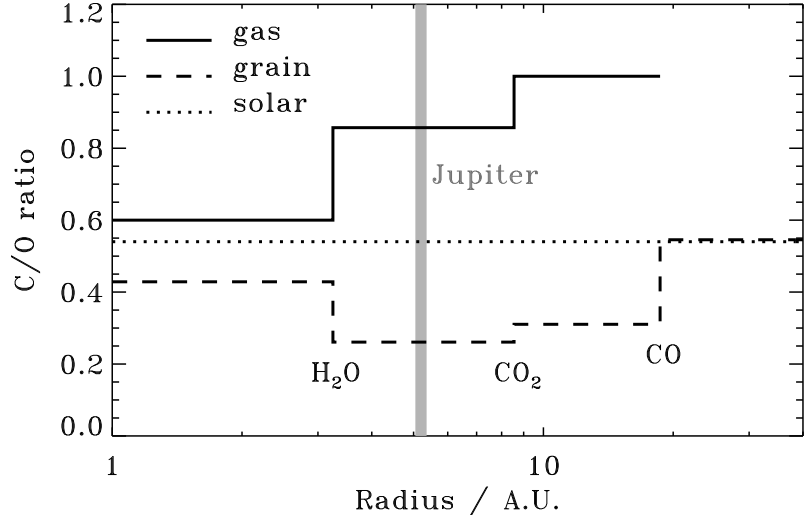} 
 \caption{The C/O ratio in the gas and in grains, assuming the temperature structure of a “typical” protoplanetary disk around a solar-type star as given by \citet{omb11}. The H$_2$O, CO$_2$, and CO snowline, along with the current location of Jupiter are marked for reference. Figure taken from \citet{omb11}.}
   \label{fig:oberg}
\end{center}
\end{figure}

The most successful connective tissue in this area relates to the C/O ratio of the gas that changes radially within the disk \citep{omb11}.   Figure~\ref{fig:oberg} illustrates how sublimation alone can change the ratio of carbon to oxygen in the ice versus gas in the midplane with stellar distance for a typical solar mass star. Beyond the CO snowline all the carbon and oxygen is frozen in the dust.  Just inside the CO snowline the ratio will be near unity at CO carries the majority of gas phase C and O because water and CO$_2$ remain frozen as ice (and so on).

In the core accretion picture, the radial position of an accreting protoplanet will determine the carbon to oxygen ratio in the core. For cores that are sufficiently massive to gravitationally capture a gas rich envelope, the carbon to oxygen ratio of the gas will tend to be the inverse of what has frozen out into solids.  Since the main phase of gas accretion is rapid ($\sim 10^4$~yrs; Chapter 6 by D'Angelo and Lissauer) the envelope C/O ratio and elemental abundances will therefore reflect the local conditions at the birth site.  This model is simplistic as it neglects additional sources or sink terms including, but not limited to pollution via planetesimal accretion \citep{Pinhas} or mixing due to core ablation \citep{Helled17}.  However, it captures the basic underpinning of the process and motivates the need to determine C/O ratios in both exoplanetary systems and in gas-rich disks.   Similar explorations can be considered for nitrogen \citep{Piso16}; however this must be tempered with the difficulty of detecting N$_2$ (a main nitrogen carrier) in disks and in exoplanet atmospheres.

Of course, the disk is evolving as Fig.~\ref{fig:oberg} presents a snapshot in time.  Snowlines will move outwards during early high accretion stages and then move inwards during later stages as the accretion rate falls off and the gaseous disk dissipates \citep[e.g.,][]{Hasegawa12,Cridland16}. These chemical effects will therefore be distributed radially as a function of the evolutionary state.  Furthermore, the ice line itself might present a trap or a favorable site for planet formation which will leave its particular chemical fingerprint \citep{Cridland16}.  The full interplay is explored in the Chapters by Mordasini and Pudritz, respectively (this volume).

The other main area to explore connections relies on the overall elemental abundance relative to hydrogen.  It is well known that Uranus and Neptune are enriched in ices; however, both Jupiter and Saturn exhibit enhanced abundances of carbon relative to solar \citep{Flasar05,wong2008}. 
In the case of Jupiter these enhancements are mirrored in numerous other elements and do not track solar abundances \citep{wong2008}.  The origin of these elemental abundance enhancements relative to solar is uncertain and has been attributed to pollution of the envelope via ice-rich planetesimals \citep{Gautier01}, early formation during the phase of radial drift \citep{Oberg16}, or perhaps dissipation of nebular hydrogen \citep{Guillot06}.  A key facet of this discussion is the disposition of oxygen in the Jovian planets which is unknown \citep{Helled14}.   For exoplanets, the question of abundance retrieval is complicated and can depend on the analysis but there are now suggestions in the literature for planets that have both superstellar and substellar abundances of C and O \citep[e.g.,][]{Lavie17,MacDonald17}.
To make connections to bulk abundances we need to trace key elemental carriers of C, O, and N within all phases of the disk lifetime while also having a reliable tracer of the hydrogen mass.  The latter point is critical and we refer the reader to a recent review on this topic by \citet{Bergin17}. 

\subsection{Global disk composition and its connection with terrestrial planet composition} 

For terrestrial worlds, in particular for the Earth, drawing connections between the chemical conditions at formation and their composition 4.6 Billion years later has a long history.  As discussed above, water is a primary condensible in planetary systems.  In the outer solar system Jovian moons and Kuiper Belt Objects often contain a significant amount of their mass as water ice \citep{Grasset17}.   In the case of the terrestrial planets, and the inner part of the asteroid belt, the water content is significantly less.  The water mass fraction of the Earth has some uncertainty, but even extreme estimates are below 0.1\% of the Earth's mass \citep{Mottl07}.  Thus it is believed that the terrestrial worlds formed interior to the snowline.

A central question is then how the Earth, and by proxy other extrasolar terrestrial worlds that formed inside their respective water snowlines, received water content (see chapter by Izidoro and Raymond, this volume). This may also be related to the presence of other ``volatile elements'' such as carbon and nitrogen (and perhaps Sulphur) that are primarily contained in compounds with volatility comparable to or below that of water (e.g. Fig.~\ref{fig:snowlines}).  Put together there are  implications for habitability writ large, and thus an important aspect is to use chemical tracers to probe the potential sources of origin for the volatile elements.
Potential source terms include in situ formation via grains with tenuous amounts of water ice, delivery by icy planetesimals (comets) or rocky bodies that contain hydrous and carbonaceous minerals (asteroids), and nebular gas capture \citep{vanDishoeck13}.  
In this regard we can isolate several lines of inquiry that are major foci of observation, theoretical, and laboratory work including: isotopic ratios and bulk composition.   Another useful line of inquiry relates to noble gases which can be used to set constraints on delivery via comets \citep{Marty17}.

\subsection{Isotopic Ratios and Chemical Fractionation}

Due to the Heisenberg Uncertainty Principle a molecule cannot exist at the bottom of the well that defines the potential of the interaction between the disparate atoms \citep[e.g.,][]{Tennyson11}.  The minimum energy is called the zero point energy, which is important for chemistry in cold regions. The ground state vibrational energy of the bond is $(1/2)h\nu_B$, where $\nu_B$ is the vibrational frequency of the bond. $\nu_B = (1/2\pi) \sqrt[]{k/\mu} $ where $k$ is the force constant and $\mu$ is the reduced mass ($\mu = m_1 \times m_2/m_1 + m_2)$ with $m_1$ and $m_1$ the mass of the atoms (in the case for a two atom molecule).
Because the force constant is invariant, this imprints a $1/\sqrt[]{\mu}$ dependence on the zero-point energy between different isotopic variants, such as H$_2$ and HD, $^{12}$CO and $^{13}$CO.  Thus, the bond is slightly stronger for the isotopically heavy molecule. This energy difference is small, but non-zero.  

Of all the stable elements hydrogen has the largest mass/energy difference, with a value of $\sim 420$~K \citep{Irikura07}. Chemical reactions that occur at temperatures well below this energy difference tend towards the stronger bond.  Hence, for a reaction with HD, the D will be preferentially transferred as opposed to the H.  This process is called fractionation and it can lead to very strong enhancements of the heavier isotope. In general isotopic enrichments of solar system material may provide a fingerprint that could provide clues regarding the formation environment of Earth's water and volatiles, but more broadly can hint at the origin of these same compounds in the disk or perhaps from the interstellar medium.  

\noindent {\em Hydrogen:}
The most well-studied isotope ratio pertains to the origins of the D/H ratio of Earth's water. The measured value is given by Vienna Standard Mean Ocean Water (VSMOW) and is D/H $=$ 1.5576 $\times 10^{-4}$. The protosolar (hydrogen) value is estimated to be $\sim 2 \times 10^{-5}$ \citep{Geiss03} and thus Earth's water has excess deuterium.  This facet is true of all measured water in interstellar bodies \citep{Cleeves14a}.  The general trend, with some heterogeneity within certain classes of objects (meteorites, Kuiper belt comets), is that the outer solar system has slightly (factor of a few) higher D/H ratios than water in the inner solar system. For a recent compilation of D/H data see \citet{Altwegg15}.  

The ultimate origin of the enhancement relates to this reaction and chemical kinetics \citep{millar_dfrac}:
\begin{equation*}
\rm{H_3^+} + HD \leftrightarrow \rm{H_2D^+} + \rm{H_2} + \Delta E,
\end{equation*}
with $\Delta$E depending on whether the reactant is ortho- or para-H$_2$.  Below $\sim$30 K, and in gas where the ortho/para ratio of H$_2$ is low \citep{flower_dfrac} this reaction can produce orders of magnitude enrichments in deuterium containing molecules (e.g. HDO, CH$_3$D, etc.).  This requires a source of ionization (either cosmic-rays or X-rays) to initiate.  Thus, the high D/H ratios cannot be created at 1 AU, as enriched water would rapidly re-equilibrate to the protosolar value \citep[or slightly higher by a factor of $\sim 3$;][]{Lecluse94}.  The source of high D/H ratios is thus either the outer disk or the pre-stellar stage \citep{Cleeves14a}, and modern models explore the question of mixing inner and outer solar system material.   

\noindent {\em Oxygen and Nitrogen:}  The isotopes of oxygen have a much smaller mass difference when compared to hydrogen and kinetic chemical effects at low temperature are minimal \citep{Langer89}. Thus the discovery of heavy oxygen isotopic enrichments in meteorites was a surprise \citep{clayton_areps}. In equilibrium, which generally pertains to geochemical systems (such as the Earth or meteoritic progenitors), isotopic chemistry has a mass dependence such that the relative abundance of $^{18}$O/$^{16}$O when plotted against $^{17}$O/$^{16}$O has a slope of 2.  This reflects the mass difference between $^{18}$O and $^{17}$O relative to $^{16}$O (e.g. $18-16/17-16 = 2$) and is consistent with isotopic ratios measured in Earth and Martian rocks.  However, oxygen isotopes in meteoritic material, when plotted in this fashion, display a slope of unity and contain an excess of the heavier oxygen isotopes.  This is not consistent with geochemistry and requires a different mechanism, which may provide clues to the origin of water in rocky bodies \citep{clayton73,clayton_areps,Thiemens06}.    A potential solution to these anomalies is photochemical self-shielding \citep{clayton_pdroxy} which has a strong isotopic signature either in the disk \citep{lyons_oxy18} or the parent cloud \citep{lbl08}.  
Heavy nitrogen isotopes are also enriched in meteoritic and cometary material \citep{Marty12}.  Here the potential origin may be linked to N$_2$ self-shielding \citep{Heays14} and/or kinetic isotopic effects \citep{Charnely02}.

\section{Astrochemical Foundations}

As we look towards a future where we have compositional information on terrestrial planet atmospheres beyond our own, we also must explore ways to understand how this composition might relate to conditions at formation.  One feasible methodology is to explore bulk chemical composition in terms of the main carriers of C and O (N remains difficult to constrain). Water stands out as a prime tracer as it is the carrier of volatile oxygen \citep{vanDishoeck13} but there is potential to explore links in the carbon inventory as well \citep{Bergin15}.  However, there are clear challenges as terrestrial worlds likely have substantial sub-surface reservoirs and one must therefore draw concurrent links to mass/radius/density \citep[e.g.,][]{Lopez14}.

At this moment we sit on the cusp of great discovery and our understanding of planet formation and its attendant effects on composition is rapidly expanding.  Today with ALMA we are constraining the conditions that exist into the giant planet forming zone. With the widely anticipated launch of the James Webb Space Telescope and the new era of large telescopes the terrestrial planet forming zone will come into greater focus.  At the same time, with the same instruments, we will be expanding our ability to characterize worlds from super-Earths, to mini-Neptunes, to Neptune-analogs, and finally massive gas giants such as Jupiter.  Now is the time where the connective tissue of formation can be explored.   First attempts at this summarized in chapters by Mordasini in particular, as well as Pudritz.
In this light, the basic astrochemical foundation discussed here remains an essential tool to provide the interpretive framework to understand the origins of planet(esimal) composition.

\bibliography{ted} 

\begin{thebibliography}{168}
\providecommand{\natexlab}[1]{#1}
\providecommand{\url}[1]{{#1}}
\providecommand{\urlprefix}{URL }
\expandafter\ifx\csname urlstyle\endcsname\relax
  \providecommand{\doi}[1]{DOI~\discretionary{}{}{}#1}\else
  \providecommand{\doi}{DOI~\discretionary{}{}{}\begingroup
  \urlstyle{rm}\Url}\fi
\providecommand{\eprint}[2][]{\url{#2}}

\bibitem[{{{\'A}d{\'a}mkovics} et~al.(2014){{\'A}d{\'a}mkovics}, {Glassgold},
  and {Najita}}]{Adamkovics14}
{{\'A}d{\'a}mkovics} M, {Glassgold} AE {Najita} JR (2014) {Shielding by Water
  and OH in FUV and X-Ray Irradiated Protoplanetary Disks}. \apj 786:135

\bibitem[{{Aikawa} et~al.(2002){Aikawa}, {van Zadelhoff}, {van Dishoeck}, and
  {Herbst}}]{aikawa_vanz02}
{Aikawa} Y, {van Zadelhoff} GJ, {van Dishoeck} EF {Herbst} E (2002) {Warm
  molecular layers in protoplanetary disks}. \aap 386:622--632

\bibitem[{{Aikawa} et~al.(2015){Aikawa}, {Furuya}, {Nomura}, and
  {Qi}}]{aikawa2015}
{Aikawa} Y, {Furuya} K, {Nomura} H {Qi} C (2015) {Analytical Formulae of
  Molecular Ion Abundances and the N$_{2}$H$^{+}$ Ring in Protoplanetary
  Disks}. \apj 807:120

\bibitem[{{Alonso-Mart{\'{\i}}nez} et~al.(2017){Alonso-Mart{\'{\i}}nez},
  {Riviere-Marichalar}, {Meeus}, {Kamp}, {Fang}, {Podio}, {Dent}, and
  {Eiroa}}]{Alonso-Martinez17}
{Alonso-Mart{\'{\i}}nez} M, {Riviere-Marichalar} P, {Meeus} G et~al. (2017)
  {Herschel GASPS spectral observations of T Tauri stars in Taurus. Unraveling
  far-infrared line emission from jets and discs}. \aap 603:A138

\bibitem[{{Altwegg} et~al.(2015){Altwegg}, {Balsiger}, {Bar-Nun}, {Berthelier},
  {Bieler}, {Bochsler}, {Briois}, {Calmonte}, {Combi}, {De Keyser},
  {Eberhardt}, {Fiethe}, {Fuselier}, {Gasc}, {Gombosi}, {Hansen}, {H{\"a}ssig},
  {J{\"a}ckel}, {Kopp}, {Korth}, {LeRoy}, {Mall}, {Marty}, {Mousis}, {Neefs},
  {Owen}, {R{\`e}me}, {Rubin}, {S{\'e}mon}, {Tzou}, {Waite}, and
  {Wurz}}]{Altwegg15}
{Altwegg} K, {Balsiger} H, {Bar-Nun} A et~al. (2015)
  {67P/Churyumov-Gerasimenko, a Jupiter family comet with a high D/H ratio}.
  Science 347(27):1261952

\bibitem[{{Anderson} et~al.(2013){Anderson}, {Bergin}, {Maret}, and
  {Wakelam}}]{Anderson13}
{Anderson} DE, {Bergin} EA, {Maret} S {Wakelam} V (2013) {New Constraints on
  the Sulfur Reservoir in the Dense Interstellar Medium Provided by Spitzer
  Observations of S I in Shocked Gas}. \apj 779:141

\bibitem[{{Andrews} et~al.(2011){Andrews}, {Wilner}, {Espaillat}, {Hughes},
  {Dullemond}, {McClure}, {Qi}, and {Brown}}]{Andrews11}
{Andrews} SM, {Wilner} DJ, {Espaillat} C et~al. (2011) {Resolved Images of
  Large Cavities in Protoplanetary Transition Disks}. \apj 732:42

\bibitem[{{Ardia} et~al.(2013){Ardia}, {Hirschmann}, {Withers}, and
  {Stanley}}]{Ardia13}
{Ardia} P, {Hirschmann} MM, {Withers} AC {Stanley} BD (2013) {Solubility of
  CH$_{4}$ in a synthetic basaltic melt, with applications to atmosphere-magma
  ocean-core partitioning of volatiles.} \gca 114:52--71

\bibitem[{{Bai} et~al.(2016){Bai}, {Ye}, {Goodman}, and {Yuan}}]{bai2016}
{Bai} XN, {Ye} J, {Goodman} J {Yuan} F (2016) {Magneto-thermal Disk Winds from
  Protoplanetary Disks}. \apj 818:152

\bibitem[{{Balbus} and {Hawley}(1991)}]{balbus91}
{Balbus} SA {Hawley} JF (1991) {A powerful local shear instability in weakly
  magnetized disks. I - Linear analysis. II - Nonlinear evolution}. \apj
  376:214--233

\bibitem[{{Bergin} et~al.(2004){Bergin}, {Calvet}, {Sitko}, {Abgrall},
  {D'Alessio}, {Herczeg}, {Roueff}, {Qi}, {Lynch}, {Russell}, {Brafford}, and
  {Perry}}]{bergin_h2}
{Bergin} E, {Calvet} N, {Sitko} ML et~al. (2004) {A New Probe of the
  Planet-forming Region in T Tauri Disks}. \apjl 614:L133--L136

\bibitem[{{Bergin} and {Williams}(2017)}]{Bergin17}
{Bergin} EA {Williams} JP (2017) {The Determination of Protoplanetary Disk
  Masses}. Formation, Evolution, and Dynamics of Young Solar Systems 445:in
  press

\bibitem[{{Bergin} et~al.(2002){Bergin}, {Alves}, {Huard}, and
  {Lada}}]{Bergin02}
{Bergin} EA, {Alves} J, {Huard} T {Lada} CJ (2002) {N$_{2}$H$^{+}$ and
  C$^{18}$O Depletion in a Cold Dark Cloud}. \apjl 570:L101--L104

\bibitem[{{Bergin} et~al.(2007){Bergin}, {Aikawa}, {Blake}, and {van
  Dishoeck}}]{bergin_ppv}
{Bergin} EA, {Aikawa} Y, {Blake} GA {van Dishoeck} EF (2007) {The Chemical
  Evolution of Protoplanetary Disks}. In: Protostars and Planets V, p 751

\bibitem[{{Bergin} et~al.(2014){Bergin}, {Cleeves}, {Crockett}, and
  {Blake}}]{Bergin14}
{Bergin} EA, {Cleeves} LI, {Crockett} N {Blake} GA (2014) Exploring the origins
  of carbon in terrestrial worlds[dagger]. Faraday Discuss 168:61--79,
  \urlprefix\url{http://dx.doi.org/10.1039/C4FD00003J}

\bibitem[{{Bergin} et~al.(2015){Bergin}, {Blake}, {Ciesla}, {Hirschmann}, and
  {Li}}]{Bergin15}
{Bergin} EA, {Blake} GA, {Ciesla} F, {Hirschmann} MM {Li} J (2015) {Tracing the
  ingredients for a habitable earth from interstellar space through planet
  formation}. Proceedings of the National Academy of Science 112:8965--8970

\bibitem[{{Bergin} et~al.(2016){Bergin}, {Du}, {Cleeves}, {Blake}, {Schwarz},
  {Visser}, and {Zhang}}]{Bergin16}
{Bergin} EA, {Du} F, {Cleeves} LI et~al. (2016) {Hydrocarbon Emission Rings in
  Protoplanetary Disks Induced by Dust Evolution}. \apj 831:101

\bibitem[{{Bethell} and {Bergin}(2009)}]{bb09}
{Bethell} T {Bergin} E (2009) {Formation and Survival of Water Vapor in the
  Terrestrial Planet-Forming Region}. Science 326:1675--

\bibitem[{{Bethell} and {Bergin}(2011{\natexlab{a}})}]{bb11b}
{Bethell} TJ {Bergin} EA (2011{\natexlab{a}}) {Photoelectric Cross-sections of
  Gas and Dust in Protoplanetary Disks}. \apj 740:7

\bibitem[{{Bethell} and {Bergin}(2011{\natexlab{b}})}]{bb11a}
{Bethell} TJ {Bergin} EA (2011{\natexlab{b}}) {The Propagation of Ly{$\alpha$}
  in Evolving Protoplanetary Disks}. \apj 739:78

\bibitem[{{B{\'e}thune} et~al.(2017){B{\'e}thune}, {Lesur}, and
  {Ferreira}}]{bethune2017}
{B{\'e}thune} W, {Lesur} G {Ferreira} J (2017) {Global simulations of
  protoplanetary disks with net magnetic flux. I. Non-ideal MHD case}. \aap
  600:A75

\bibitem[{{Blandford} and {Payne}(1982)}]{blandford1982}
{Blandford} RD {Payne} DG (1982) {Hydromagnetic flows from accretion discs and
  the production of radio jets}. \mnras 199:883--903

\bibitem[{{Blevins} et~al.(2016){Blevins}, {Pontoppidan}, {Banzatti}, {Zhang},
  {Najita}, {Carr}, {Salyk}, and {Blake}}]{Blevins16}
{Blevins} SM, {Pontoppidan} KM, {Banzatti} A et~al. (2016) {Measurements of
  water surface snow lines in classical protoplanetary disks}. \apj p in press.

\bibitem[{{Calvet} et~al.(1991){Calvet}, {Patino}, {Magris}, and
  {D'Alessio}}]{calvet91}
{Calvet} N, {Patino} A, {Magris} GC {D'Alessio} P (1991) {Irradiation of
  accretion disks around young objects. I - Near-infrared CO bands}. \apj
  380:617--630

\bibitem[{{Cazaux} and {Tielens}(2002)}]{Cazaux02}
{Cazaux} S {Tielens} AGGM (2002) {Molecular Hydrogen Formation in the
  Interstellar Medium}. \apjl 575:L29--L32

\bibitem[{{Cazaux} and {Tielens}(2004)}]{ct04}
{Cazaux} S {Tielens} AGGM (2004) {H$_{2}$ Formation on Grain Surfaces}. \apj
  604:222--237

\bibitem[{{Cazaux} et~al.(2005){Cazaux}, {Caselli}, {Tielens}, {LeBourlot}, and
  {Walmsley}}]{Cazaux05}
{Cazaux} S, {Caselli} P, {Tielens} AGGM, {LeBourlot} J {Walmsley} M (2005)
  {Molecular Hydrogen formation on grain surfaces}. In: {Saija} R
  {Cecchi-Pestellini} C (eds) Journal of Physics Conference Series, Journal of
  Physics Conference Series, vol~6, pp 155--160,
  \doi{10.1088/1742-6596/6/1/016}

\bibitem[{{Charnley}(1997)}]{charnley1997}
{Charnley} SB (1997) {Chemical models of interstellar gas-grain processes. III
  - Molecular depletion in NGC 2024}. \mnras 291:455

\bibitem[{{Charnley} and {Rodgers}(2002)}]{Charnely02}
{Charnley} SB {Rodgers} SD (2002) {The End of Interstellar Chemistry as the
  Origin of Nitrogen in Comets and Meteorites}. \apjl 569:L133--L137

\bibitem[{{Chiang} and {Goldreich}(1997)}]{cg97}
{Chiang} EI {Goldreich} P (1997) {Spectral Energy Distributions of T Tauri
  Stars with Passive Circumstellar Disks}. \apj 490:368--+

\bibitem[{{Clayton}(1993)}]{clayton_areps}
{Clayton} RN (1993) {Oxygen isotopes in meteorites}. Annual Review of Earth and
  Planetary Sciences 21:115--149

\bibitem[{{Clayton}(2002)}]{clayton_pdroxy}
{Clayton} RN (2002) {Solar System: Self-shielding in the solar nebula}. \nat
  415:860--861

\bibitem[{{Clayton} et~al.(1973){Clayton}, {Grossman}, and
  {Mayeda}}]{clayton73}
{Clayton} RN, {Grossman} L {Mayeda} TK (1973) {A Component of Primitive Nuclear
  Composition in Carbonaceous Meteorites}. Science 182:485--488

\bibitem[{{Cleeves}(2016)}]{cleeves2016}
{Cleeves} LI (2016) {Multiple Carbon Monoxide Snow Lines in Disks Sculpted by
  Radial Drift}. \apjl 816:L21

\bibitem[{{Cleeves} et~al.(2013){Cleeves}, {Adams}, and {Bergin}}]{cleeves13a}
{Cleeves} LI, {Adams} FC {Bergin} EA (2013) {Exclusion of Cosmic Rays in
  Protoplanetary Disks: Stellar and Magnetic Effects}. \apj 772:5

\bibitem[{{Cleeves} et~al.(2014){Cleeves}, {Bergin}, {Alexander}, {Du},
  {Graninger}, {{\"O}berg}, and {Harries}}]{Cleeves14a}
{Cleeves} LI, {Bergin} EA, {Alexander} CMO et~al. (2014) {The ancient heritage
  of water ice in the solar system}. Science 345:1590--1593

\bibitem[{{Cleeves} et~al.(2015){Cleeves}, {Bergin}, {Qi}, {Adams}, and
  {{\"O}berg}}]{Cleeves15}
{Cleeves} LI, {Bergin} EA, {Qi} C, {Adams} FC {{\"O}berg} KI (2015)
  {Constraining the X-Ray and Cosmic-Ray Ionization Chemistry of the TW Hya
  Protoplanetary Disk: Evidence for a Sub-interstellar Cosmic-Ray Rate}. \apj
  799:204

\bibitem[{{Cridland} et~al.(2016){Cridland}, {Pudritz}, and
  {Birnstiel}}]{Cridland16}
{Cridland} AJ, {Pudritz} RE {Birnstiel} T (2016) {Radial Drift of Dust in
  Protoplanetary Disks: The Evolution of Ice lines and Dead zones}. MNRAS p
  submitted

\bibitem[{{D'Alessio} et~al.(1999){D'Alessio}, {Calvet}, {Hartmann}, {Lizano},
  and {Cant{\'o}}}]{dalessio1999}
{D'Alessio} P, {Calvet} N, {Hartmann} L, {Lizano} S {Cant{\'o}} J (1999)
  {Accretion Disks around Young Objects. II. Tests of Well-mixed Models with
  ISM Dust}. \apj 527:893--909

\bibitem[{{D'Alessio} et~al.(2001){D'Alessio}, {Calvet}, and
  {Hartmann}}]{dalessio01}
{D'Alessio} P, {Calvet} N {Hartmann} L (2001) {Accretion Disks around Young
  Objects. III. Grain Growth}. \apj 553:321--334

\bibitem[{{D'Alessio} et~al.(2005){D'Alessio}, {Calvet}, and {Woolum}}]{dcw05}
{D'Alessio} P, {Calvet} N {Woolum} DS (2005) {Thermal Structure of
  Protoplanetary Disks}. In: {Krot} AN, {Scott} ERD {Reipurth} B (eds)
  Chondrites and the Protoplanetary Disk, Astronomical Society of the Pacific
  Conference Series, vol 341, pp 353--+

\bibitem[{{D'Alessio} et~al.(2006){D'Alessio}, {Calvet}, {Hartmann},
  {Franco-Hern{\'a}ndez}, and {Serv{\'{\i}}n}}]{dalessio06}
{D'Alessio} P, {Calvet} N, {Hartmann} L, {Franco-Hern{\'a}ndez} R
  {Serv{\'{\i}}n} H (2006) {Effects of Dust Growth and Settling in T Tauri
  Disks}. \apj 638:314--335

\bibitem[{{Dalgarno}(2006)}]{Dalgarno06}
{Dalgarno} A (2006) {Interstellar Chemistry Special Feature: The galactic
  cosmic ray ionization rate}. Proceedings of the National Academy of Science
  103:12,269--12,273

\bibitem[{{Debes} et~al.(2013){Debes}, {Jang-Condell}, {Weinberger}, {Roberge},
  and {Schneider}}]{Debes13}
{Debes} JH, {Jang-Condell} H, {Weinberger} AJ, {Roberge} A {Schneider} G (2013)
  {The 0.5-2.22 {$\mu$}m Scattered Light Spectrum of the Disk around TW Hya:
  Detection of a Partially Filled Disk Gap at 80 AU}. \apj 771:45

\bibitem[{{Dominik} et~al.(2005){Dominik}, {Ceccarelli}, {Hollenbach}, and
  {Kaufman}}]{dchk05}
{Dominik} C, {Ceccarelli} C, {Hollenbach} D {Kaufman} M (2005) {Gas-Phase Water
  in the Surface Layer of Protoplanetary Disks}. \apjl 635:L85--L88

\bibitem[{{Druard} and {Wakelam}(2012)}]{Druard12}
{Druard} C {Wakelam} V (2012) {Polysulphanes on interstellar grains as a
  possible reservoir of interstellar sulphur}. \mnras 426:354--359

\bibitem[{{Du} and {Bergin}(2014)}]{Du14}
{Du} F {Bergin} EA (2014) {Water Vapor Distribution in Protoplanetary Disks}.
  \apj 792:2

\bibitem[{{Du} et~al.(2015){Du}, {Bergin}, and {Hogerheijde}}]{Du15}
{Du} F, {Bergin} EA {Hogerheijde} MR (2015) {Volatile depletion in the TW
  Hydrae disk atmosphere}. \apjl 807:L32

\bibitem[{{Du} et~al.(2017){Du}, {Bergin}, {Hogerheijde}, {van Dishoeck},
  {Blake}, {Bruderer}, {Cleeves}, {Dominik}, {Fedele}, {Lis}, {Melnick},
  {Neufeld}, {Pearson}, and {Yildiz}}]{du2017}
{Du} F, {Bergin} EA, {Hogerheijde} M et~al. (2017) {Survey of Cold Water Lines
  in Protoplanetary Disks: Indications of Systematic Volatile Depletion}. \apj
  842:98

\bibitem[{{Dullemond} and {Dominik}(2004)}]{dd04}
{Dullemond} CP {Dominik} C (2004) {The effect of dust settling on the
  appearance of protoplanetary disks}. \aap 421:1075--1086

\bibitem[{{Dutrey} et~al.(2014){Dutrey}, {Semenov}, {Chapillon}, {Gorti},
  {Guilloteau}, {Hersant}, {Hogerheijde}, {Hughes}, {Meeus}, {Nomura},
  {Pi{\'e}tu}, {Qi}, and {Wakelam}}]{Dutrey14}
{Dutrey} A, {Semenov} D, {Chapillon} E et~al. (2014) {Physical and Chemical
  Structure of Planet-Forming Disks Probed by Millimeter Observations and
  Modeling}. Protostars and Planets VI pp 317--338

\bibitem[{{Dutrey} et~al.(2017){Dutrey}, {Guilloteau}, {Pi{\'e}tu},
  {Chapillon}, {Wakelam}, {Di Folco}, {Stoecklin}, {Denis-Alpizar}, {Gorti},
  {Teague}, {Henning}, {Semenov}, and {Grosso}}]{dutrey2017}
{Dutrey} A, {Guilloteau} S, {Pi{\'e}tu} V et~al. (2017) {The Flying Saucer:
  Tomography of the thermal and density gas structure of an edge-on
  protoplanetary disk}. ArXiv e-prints

\bibitem[{{Facchini} et~al.(2017){Facchini}, {Birnstiel}, {Bruderer}, and {van
  Dishoeck}}]{facchini2017}
{Facchini} S, {Birnstiel} T, {Bruderer} S {van Dishoeck} EF (2017) {Different
  dust and gas radial extents in protoplanetary disks: consistent models of
  grain growth and CO emission}. \aap 605:A16

\bibitem[{{Fayolle} et~al.(2016){Fayolle}, {Balfe}, {Loomis}, {Bergner},
  {Graninger}, {Rajappan}, and {{\"O}berg}}]{fayolle16}
{Fayolle} EC, {Balfe} J, {Loomis} R et~al. (2016) {N$_{2}$ and CO Desorption
  Energies from Water Ice}. \apjl 816:L28

\bibitem[{{Flaherty} et~al.(2015){Flaherty}, {Hughes}, {Rosenfeld}, {Andrews},
  {Chiang}, {Simon}, {Kerzner}, and {Wilner}}]{Flaherty15}
{Flaherty} KM, {Hughes} AM, {Rosenfeld} KA et~al. (2015) {Weak Turbulence in
  the HD 163296 Protoplanetary Disk Revealed by ALMA CO Observations}. \apj
  813:99

\bibitem[{{Flasar} et~al.(2005){Flasar}, {Achterberg}, {Conrath}, {Gierasch},
  {Kunde}, {Nixon}, {Bjoraker}, {Jennings}, {Romani}, {Simon-Miller},
  {B{\'e}zard}, {Coustenis}, {Irwin}, {Teanby}, {Brasunas}, {Pearl}, {Segura},
  {Carlson}, {Mamoutkine}, {Schinder}, {Barucci}, {Courtin}, {Fouchet},
  {Gautier}, {Lellouch}, {Marten}, {Prang{\'e}}, {Vinatier}, {Strobel},
  {Calcutt}, {Read}, {Taylor}, {Bowles}, {Samuelson}, {Orton}, {Spilker},
  {Owen}, {Spencer}, {Showalter}, {Ferrari}, {Abbas}, {Raulin}, {Edgington},
  {Ade}, and {Wishnow}}]{Flasar05}
{Flasar} FM, {Achterberg} RK, {Conrath} BJ et~al. (2005) {Titan's Atmospheric
  Temperatures, Winds, and Composition}. Science 308:975--978

\bibitem[{{Flower} et~al.(2004){Flower}, {Pineau des For{\^e}ts}, and
  {Walmsley}}]{flower_dfrac}
{Flower} DR, {Pineau des For{\^e}ts} G {Walmsley} CM (2004)
  {Multiply-deuterated species in prestellar cores}. \aap 427:887--893

\bibitem[{{Fogel} et~al.(2011){Fogel}, {Bethell}, {Bergin}, {Calvet}, and
  {Semenov}}]{fogel11}
{Fogel} JKJ, {Bethell} TJ, {Bergin} EA, {Calvet} N {Semenov} D (2011)
  {Chemistry of a Protoplanetary Disk with Grain Settling and Ly{$\alpha$}
  Radiation}. \apj 726:29

\bibitem[{{France} et~al.(2012){France}, {Schindhelm}, {Herczeg}, {Brown},
  {Abgrall}, {Alexander}, {Bergin}, {Brown}, {Linsky}, {Roueff}, and
  {Yang}}]{France12}
{France} K, {Schindhelm} E, {Herczeg} GJ et~al. (2012) {A Hubble Space
  Telescope Survey of H$_{2}$ Emission in the Circumstellar Environments of
  Young Stars}. \apj 756:171

\bibitem[{{France} et~al.(2014){France}, {Herczeg}, {McJunkin}, and
  {Penton}}]{France14}
{France} K, {Herczeg} GJ, {McJunkin} M {Penton} SV (2014) {CO/H$_{2}$ Abundance
  Ratio $\sim$10$^{-4}$ in a Protoplanetary Disk}. \apj 794:160

\bibitem[{{Fraser} et~al.(2001){Fraser}, {Collings}, {McCoustra}, and
  {Williams}}]{fraser_h2obind}
{Fraser} HJ, {Collings} MP, {McCoustra} MRS {Williams} DA (2001) {Thermal
  desorption of water ice in the interstellar medium}. \mnras 327:1165--1172

\bibitem[{{Furlan} et~al.(2005)}]{furlan05}
{Furlan} E et~al. (2005) {Colors of Classical T Tauri Stars in Taurus Derived
  from Spitzer Infrared Spectrograph Spectra: Indication of Dust Settling}.
  \apjl 628:L65--L68

\bibitem[{{Furuya} and {Aikawa}(2014)}]{Furuya14}
{Furuya} K {Aikawa} Y (2014) {Reprocessing of Ices in Turbulent Protoplanetary
  Disks: Carbon and Nitrogen Chemistry}. \apj 790:97

\bibitem[{{Gautier} et~al.(2001){Gautier}, {Hersant}, {Mousis}, and
  {Lunine}}]{Gautier01}
{Gautier} D, {Hersant} F, {Mousis} O {Lunine} JI (2001) {Enrichments in
  Volatiles in Jupiter: A New Interpretation of the Galileo Measurements}.
  \apjl 550:L227--L230

\bibitem[{{Geiss} and {Gloeckler}(2003)}]{Geiss03}
{Geiss} J {Gloeckler} G (2003) {Isotopic Composition of H, HE and NE in the
  Protosolar Cloud}. \ssr 106:3--18

\bibitem[{{Gorti} and {Hollenbach}(2004)}]{gortihollenbach04}
{Gorti} U {Hollenbach} D (2004) {Models of Chemistry, Thermal Balance, and
  Infrared Spectra from Intermediate-Aged Disks around G and K Stars}. \apj
  613:424--447

\bibitem[{{Grasset} et~al.(2017){Grasset}, {Castillo-Rogez}, {Guillot},
  {Fletcher}, and {Tosi}}]{Grasset17}
{Grasset} O, {Castillo-Rogez} J, {Guillot} T, {Fletcher} LN {Tosi} F (2017)
  {Water and Volatiles in the Outer Solar System}. \ssr

\bibitem[{{G{\"u}del}(2004)}]{Guedel04}
{G{\"u}del} M (2004) {X-ray astronomy of stellar coronae}. \aapr 12:71--237

\bibitem[{{Guillot} and {Hueso}(2006)}]{Guillot06}
{Guillot} T {Hueso} R (2006) {The composition of Jupiter: sign of a
  (relatively) late formation in a chemically evolved protosolar disc}. \mnras
  367:L47--L51

\bibitem[{{Gullbring} et~al.(2000){Gullbring}, {Calvet}, {Muzerolle}, and
  {Hartmann}}]{gullbring00}
{Gullbring} E, {Calvet} N, {Muzerolle} J {Hartmann} L (2000) {The Structure and
  Emission of the Accretion Shock in T Tauri Stars. II. The
  Ultraviolet-Continuum Emission}. \apj 544:927--932

\bibitem[{{Habing}(1968)}]{habing68}
{Habing} HJ (1968) {The interstellar radiation density between 912 A and 2400
  A}. \bain 19:421--+

\bibitem[{{Hartmann} et~al.(1998){Hartmann}, {Calvet}, {Gullbring}, and
  {D'Alessio}}]{hartmann1998}
{Hartmann} L, {Calvet} N, {Gullbring} E {D'Alessio} P (1998) {Accretion and the
  Evolution of T Tauri Disks}. \apj 495:385--400

\bibitem[{{Hasegawa} and {Pudritz}(2012)}]{Hasegawa12}
{Hasegawa} Y {Pudritz} RE (2012) {Evolutionary Tracks of Trapped, Accreting
  Protoplanets: The Origin of the Observed Mass-Period Relation}. \apj 760:117

\bibitem[{{Hayashi}(1981)}]{hayashi_mmsn}
{Hayashi} C (1981) {Structure of the Solar Nebula, Growth and Decay of Magnetic
  Fields and Effects of Magnetic and Turbulent Viscosities on the Nebula}.
  Progress of Theoretical Physics Supplement 70:35--53

\bibitem[{{Heays} et~al.(2014){Heays}, {Visser}, {Gredel}, {Ubachs}, {Lewis},
  {Gibson}, and {van Dishoeck}}]{Heays14}
{Heays} AN, {Visser} R, {Gredel} R et~al. (2014) {Isotope selective
  photodissociation of N$_{2}$ by the interstellar radiation field and cosmic
  rays}. \aap 562:A61

\bibitem[{{Helled} and {Guillot}(2017)}]{Helled17}
{Helled} R {Guillot} T (2017) {Internal Structure of Giant and Icy Planets:
  Importance of Heavy Elements and Mixing}. ArXiv e-prints

\bibitem[{{Helled} and {Lunine}(2014)}]{Helled14}
{Helled} R {Lunine} J (2014) {Measuring Jupiter's water abundance by Juno: the
  link between interior and formation models}. \mnras 441:2273--2279

\bibitem[{{Henke} et~al.(1993){Henke}, {Gullikson}, and {Davis}}]{Henke93}
{Henke} BL, {Gullikson} EM {Davis} JC (1993) {X-Ray Interactions:
  Photoabsorption, Scattering, Transmission, and Reflection at E = 50-30,000
  eV, Z = 1-92}. Atomic Data and Nuclear Data Tables 54:181--342

\bibitem[{{Henning} and {Semenov}(2013)}]{henning13}
{Henning} T {Semenov} D (2013) {Chemistry in Protoplanetary Disks}. ArXiv
  e-prints

\bibitem[{{Herbst} and {Klemperer}(1973)}]{hk73}
{Herbst} E {Klemperer} W (1973) {The Formation and Depletion of Molecules in
  Dense Interstellar Clouds}. \apj 185:505--534

\bibitem[{{Herczeg} et~al.(2002){Herczeg}, {Linsky}, {Valenti}, {Johns-Krull},
  and {Wood}}]{herczeg_twhya1}
{Herczeg} GJ, {Linsky} JL, {Valenti} JA, {Johns-Krull} CM {Wood} BE (2002) {The
  Far-Ultraviolet Spectrum of TW Hydrae. I. Observations of H$_{2}$
  Fluorescence}. \apj 572:310--325

\bibitem[{{Herrero} et~al.(2010){Herrero}, {G{\'a}lvez}, {Mat{\'e}}, and
  {Escribano}}]{Herrero10}
{Herrero} VJ, {G{\'a}lvez} {\'O}, {Mat{\'e}} B {Escribano} R (2010)
  {Interaction of CH4 and H2O in ice mixtures}. Physical Chemistry Chemical
  Physics (Incorporating Faraday Transactions) 12:3164

\bibitem[{{Hogerheijde} et~al.(2011){Hogerheijde}, {Bergin}, {Brinch}
  et~al.}]{hoger11a}
{Hogerheijde} MR, {Bergin} EA, {Brinch} C et~al. (2011) {Detection of the Water
  Reservoir in a Forming Planetary System}. Science 334:338--340

\bibitem[{{Hollenbach} et~al.(2009){Hollenbach}, {Kaufman}, {Bergin}, and
  {Melnick}}]{hkbm09}
{Hollenbach} D, {Kaufman} MJ, {Bergin} EA {Melnick} GJ (2009) {Water, O$_{2}$,
  and Ice in Molecular Clouds}. \apj 690:1497--1521

\bibitem[{{Hollenbach} and {Tielens}(1999)}]{ht_rvmp}
{Hollenbach} DJ {Tielens} AGGM (1999) {Photodissociation regions in the
  interstellar medium of galaxies}. Reviews of Modern Physics 71:173--230

\bibitem[{{Howard} et~al.(2013){Howard}, {Sandell}, {Vacca}, {Duch{\^e}ne},
  {Mathews}, {Augereau}, {Barrado}, {Dent}, {Eiroa}, {Grady}, {Kamp}, {Meeus},
  {M{\'e}nard}, {Pinte}, {Podio}, {Riviere-Marichalar}, {Roberge}, {Thi},
  {Vicente}, and {Williams}}]{Howard13}
{Howard} CD, {Sandell} G, {Vacca} WD et~al. (2013) {Herschel/PACS Survey of
  Protoplanetary Disks in Taurus/Auriga Observations of [O I] and [C II], and
  Far-infrared Continuum}. \apj 776:21

\bibitem[{{Hughes} et~al.(2011){Hughes}, {Wilner}, {Andrews}, {Qi}, and
  {Hogerheijde}}]{hughes2011}
{Hughes} AM, {Wilner} DJ, {Andrews} SM, {Qi} C {Hogerheijde} MR (2011)
  {Empirical Constraints on Turbulence in Protoplanetary Accretion Disks}. \apj
  727:85

\bibitem[{{Igea} and {Glassgold}(1999)}]{ig99}
{Igea} J {Glassgold} AE (1999) {X-Ray Ionization of the Disks of Young Stellar
  Objects}. \apj 518:848--858

\bibitem[{{Ilgner} and {Nelson}(2006)}]{ilgner_diskmetals}
{Ilgner} M {Nelson} RP (2006) {On the ionisation fraction in protoplanetary
  disks. II. The effect of turbulent mixing on gas-phase chemistry}. \aap
  445:223--232

\bibitem[{{Irikura}(2007)}]{Irikura07}
{Irikura} KK (2007) {Experimental Vibrational Zero-Point Energies: Diatomic
  Molecules}. Journal of Physical and Chemical Reference Data 36:389--397

\bibitem[{{Isella} et~al.(2009){Isella}, {Carpenter}, and {Sargent}}]{Isella09}
{Isella} A, {Carpenter} JM {Sargent} AI (2009) {Structure and Evolution of
  Pre-main-sequence Circumstellar Disks}. \apj 701:260--282

\bibitem[{{Kama} et~al.(2016){Kama}, {Bruderer}, {van Dishoeck}, {Hogerheijde},
  {Folsom}, {Miotello}, {Fedele}, {Belloche}, {G{\"u}sten}, and
  {Wyrowski}}]{Kama16a}
{Kama} M, {Bruderer} S, {van Dishoeck} EF et~al. (2016) {Volatile carbon
  locking and release in protoplanetary disks. A study of TW Hya and HD
  100546}. \aap p in press.

\bibitem[{{Kamp} and {Dullemond}(2004)}]{Kamp04}
{Kamp} I {Dullemond} CP (2004) {The Gas Temperature in the Surface Layers of
  Protoplanetary Disks}. \apj 615:991--999

\bibitem[{{Kamp} et~al.(2013){Kamp}, {Thi}, {Meeus}, {Woitke}, {Pinte},
  {Meijerink}, {Spaans}, {Pascucci}, {Aresu}, and {Dent}}]{Kamp13}
{Kamp} I, {Thi} WF, {Meeus} G et~al. (2013) {Uncertainties in water chemistry
  in disks: An application to TW Hydrae}. \aap 559:A24

\bibitem[{{Kastner} et~al.(2015){Kastner}, {Qi}, {Gorti}, {Hily-Blant},
  {Oberg}, {Forveille}, {Andrews}, and {Wilner}}]{Kastner15}
{Kastner} JH, {Qi} C, {Gorti} U et~al. (2015) {A Ring of C$_{2}$H in the
  Molecular Disk Orbiting TW Hya}. \apj 806:75

\bibitem[{{Kaufman} and {Neufeld}(1996)}]{kn96}
{Kaufman} MJ {Neufeld} DA (1996) {Water Maser Emission from Magnetohydrodynamic
  Shock Waves}. \apj 456:250--+

\bibitem[{{Kenyon} and {Hartmann}(1995)}]{kh95}
{Kenyon} SJ {Hartmann} L (1995) {Pre-Main-Sequence Evolution in the
  Taurus-Auriga Molecular Cloud}. \apjs 101:117--+

\bibitem[{{Konigl} and {Pudritz}(2000)}]{konigl2000}
{Konigl} A {Pudritz} RE (2000) {Disk Winds and the Accretion-Outflow
  Connection}. Protostars and Planets IV p 759

\bibitem[{{Krijt} et~al.(2016){Krijt}, {Ciesla}, and {Bergin}}]{Krijt16}
{Krijt} S, {Ciesla} FJ {Bergin} EA (2016) {Tracing Water Vapor and Ice During
  Dust Growth}. \apj p in press

\bibitem[{{Langer} and {Graedel}(1989)}]{Langer89}
{Langer} WD {Graedel} TE (1989) {Ion-molecule chemistry of dense interstellar
  clouds - Nitrogen-, oxygen-, and carbon-bearing molecule abundances and
  isotopic ratios}. \apjs 69:241--269

\bibitem[{{Lavie} et~al.(2017){Lavie}, {Mendon{\c c}a}, {Mordasini}, {Malik},
  {Bonnefoy}, {Demory}, {Oreshenko}, {Grimm}, {Ehrenreich}, and
  {Heng}}]{Lavie17}
{Lavie} B, {Mendon{\c c}a} JM, {Mordasini} C et~al. (2017) {HELIOS-RETRIEVAL:
  An Open-source, Nested Sampling Atmospheric Retrieval Code; Application to
  the HR 8799 Exoplanets and Inferred Constraints for Planet Formation}. \aj
  154:91

\bibitem[{{L{\'e}cluse} and {Robert}(1994)}]{Lecluse94}
{L{\'e}cluse} C {Robert} F (1994) {Hydrogen isotope exchange reaction rates:
  Origin of water in the inner solar system}. \gca 58:2927--2939

\bibitem[{{Lee} et~al.(2008){Lee}, {Bergin}, and {Lyons}}]{lbl08}
{Lee} JE, {Bergin} EA {Lyons} JR (2008) {Oxygen isotope anomalies of the Sun
  and the original environment of the solar system}. Meteoritics and Planetary
  Science 43:1351--1362

\bibitem[{{Loomis} et~al.(2015){Loomis}, {Cleeves}, {{\"O}berg}, {Guzman}, and
  {Andrews}}]{loomis2015}
{Loomis} RA, {Cleeves} LI, {{\"O}berg} KI, {Guzman} VV {Andrews} SM (2015) {The
  Distribution and Chemistry of H$_{2}$CO in the DM Tau Protoplanetary Disk}.
  \apjl 809:L25

\bibitem[{{Lopez} and {Fortney}(2014)}]{Lopez14}
{Lopez} ED {Fortney} JJ (2014) {Understanding the Mass-Radius Relation for
  Sub-neptunes: Radius as a Proxy for Composition}. \apj 792:1

\bibitem[{{Lyons} and {Young}(2005)}]{lyons_oxy18}
{Lyons} JR {Young} ED (2005) {CO self-shielding as the origin of oxygen isotope
  anomalies in the early solar nebula}. \nat 435:317--320

\bibitem[{{MacDonald} and {Madhusudhan}(2017)}]{MacDonald17}
{MacDonald} RJ {Madhusudhan} N (2017) {HD 209458b in new light: evidence of
  nitrogen chemistry, patchy clouds and sub-solar water}. \mnras 469:1979--1996

\bibitem[{{Maret} and {Bergin}(2007)}]{mb07}
{Maret} S {Bergin} EA (2007) {The Ionization Fraction of Barnard 68:
  Implications for Star and Planet Formation}. \apj 664:956

\bibitem[{{Mart{\'{\i}}n-Dom{\'e}nech}
  et~al.(2014){Mart{\'{\i}}n-Dom{\'e}nech}, {Mu{\~n}oz Caro}, {Bueno}, and
  {Goesmann}}]{Martin-Domenech14}
{Mart{\'{\i}}n-Dom{\'e}nech} R, {Mu{\~n}oz Caro} GM, {Bueno} J {Goesmann} F
  (2014) {Thermal desorption of circumstellar and cometary ice analogs}. \aap
  564:A8

\bibitem[{{Marty}(2012)}]{Marty12}
{Marty} B (2012) {The origins and concentrations of water, carbon, nitrogen and
  noble gases on Earth}. Earth and Planetary Science Letters 313:56--66

\bibitem[{{Marty} et~al.(2017){Marty}, {Altwegg}, {Balsiger}, {Bar-Nun},
  {Bekaert}, {Berthelier}, {Bieler}, {Briois}, {Calmonte}, {Combi}, {De
  Keyser}, {Fiethe}, {Fuselier}, {Gasc}, {Gombosi}, {Hansen}, {H{\"a}ssig},
  {J{\"a}ckel}, {Kopp}, {Korth}, {Le Roy}, {Mall}, {Mousis}, {Owen},
  {R{\`e}me}, {Rubin}, {S{\'e}mon}, {Tzou}, {Waite}, and {Wurz}}]{Marty17}
{Marty} B, {Altwegg} K, {Balsiger} H et~al. (2017) {Xenon isotopes in
  67P/Churyumov-Gerasimenko show that comets contributed to Earth's
  atmosphere}. Science 356:1069--1072

\bibitem[{{Millar} et~al.(1989){Millar}, {Bennett}, and
  {Herbst}}]{millar_dfrac}
{Millar} TJ, {Bennett} A {Herbst} E (1989) {Deuterium fractionation in dense
  interstellar clouds}. \apj 340:906--920

\bibitem[{{Morrison} and {McCammon}(1983)}]{mmc83}
{Morrison} R {McCammon} D (1983) {Interstellar photoelectric absorption cross
  sections, 0.03-10 keV}. \apj 270:119--122

\bibitem[{{Moses}(2014)}]{Moses14}
{Moses} JI (2014) {Chemical kinetics on extrasolar planets}. Philosophical
  Transactions of the Royal Society of London Series A 372:20130,073--20130,073

\bibitem[{{Mottl} et~al.(2007){Mottl}, {Glazer}, {Kaiser}, and
  {Meech}}]{Mottl07}
{Mottl} M, {Glazer} B, {Kaiser} R {Meech} K (2007) {Water and astrobiology}.
  Chemie der Erde / Geochemistry 67:253--282

\bibitem[{{Musiolik} et~al.(2016{\natexlab{a}}){Musiolik}, {Teiser},
  {Jankowski}, and {Wurm}}]{musiolik2016}
{Musiolik} G, {Teiser} J, {Jankowski} T {Wurm} G (2016{\natexlab{a}})
  {Collisions of CO$_{2}$ Ice Grains in Planet Formation}. \apj 818:16

\bibitem[{{Musiolik} et~al.(2016{\natexlab{b}}){Musiolik}, {Teiser},
  {Jankowski}, and {Wurm}}]{musiolik2016b}
{Musiolik} G, {Teiser} J, {Jankowski} T {Wurm} G (2016{\natexlab{b}}) {Ice
  Grain Collisions in Comparison: CO2, H2O, and Their Mixtures}. \apj 827:63

\bibitem[{{Najita} et~al.(2011){Najita}, {{\'A}d{\'a}mkovics}, and
  {Glassgold}}]{Najita11}
{Najita} JR, {{\'A}d{\'a}mkovics} M {Glassgold} AE (2011) {Formation of Organic
  Molecules and Water in Warm Disk Atmospheres}. \apj 743:147

\bibitem[{{Nomura} and {Millar}(2005)}]{Nomura05}
{Nomura} H {Millar} TJ (2005) {Molecular hydrogen emission from protoplanetary
  disks}. \aap 438:923--938

\bibitem[{{{\"O}berg} and {Bergin}(2016)}]{Oberg16}
{{\"O}berg} KI {Bergin} EA (2016) {Excess C/O and C/H in Outer Protoplanetary
  Disk Gas}. \apjl 831:L19

\bibitem[{{{\"O}berg} et~al.(2009){{\"O}berg}, {van Dishoeck}, and
  {Linnartz}}]{Oberg09}
{{\"O}berg} KI, {van Dishoeck} EF {Linnartz} H (2009) {Photodesorption of ices
  I: CO, N$_{2}$, and CO$_{2}$}. \aap 496:281--293

\bibitem[{{{\"O}berg} et~al.(2011){{\"O}berg}, {Murray-Clay}, and
  {Bergin}}]{omb11}
{{\"O}berg} KI, {Murray-Clay} R {Bergin} EA (2011) {The Effects of Snowlines on
  C/O in Planetary Atmospheres}. \apjl 743:L16

\bibitem[{{{\"O}berg} et~al.(2015){{\"O}berg}, {Furuya}, {Loomis}, {Aikawa},
  {Andrews}, {Qi}, {van Dishoeck}, and {Wilner}}]{Oberg15a}
{{\"O}berg} KI, {Furuya} K, {Loomis} R et~al. (2015) {Double DCO$^{+}$ Rings
  Reveal CO Ice Desorption in the Outer Disk Around IM Lup}. \apj 810:112

\bibitem[{{Padovani} et~al.(2009){Padovani}, {Galli}, and
  {Glassgold}}]{Padovani09}
{Padovani} M, {Galli} D {Glassgold} AE (2009) {Cosmic-ray ionization of
  molecular clouds}. \aap 501:619--631

\bibitem[{{Perez-Becker} and {Chiang}(2011)}]{Perez-Becker11}
{Perez-Becker} D {Chiang} E (2011) {Surface Layer Accretion in Conventional and
  Transitional Disks Driven by Far-ultraviolet Ionization}. \apj 735:8

\bibitem[{{Pinhas} et~al.(2016){Pinhas}, {Madhusudhan}, and {Clarke}}]{Pinhas}
{Pinhas} A, {Madhusudhan} N {Clarke} C (2016) {Efficiency of planetesimal
  ablation in giant planetary envelopes}. \mnras 463:4516--4532

\bibitem[{{Pinte} et~al.(2016){Pinte}, {Dent}, {M{\'e}nard}, {Hales}, {Hill},
  {Cortes}, and {de Gregorio-Monsalvo}}]{Pinte16}
{Pinte} C, {Dent} WRF, {M{\'e}nard} F et~al. (2016) {Dust and Gas in the Disk
  of HL Tauri: Surface Density, Dust Settling, and Dust-to-gas Ratio}. \apj
  816:25

\bibitem[{{Piso} et~al.(2016){Piso}, {Pegues}, and {{\"O}berg}}]{Piso16}
{Piso} AMA, {Pegues} J {{\"O}berg} KI (2016) {The Role of Ice Compositions for
  Snowlines and the C/N/O Ratios in Active Disks}. \apj 833:203

\bibitem[{{Pontoppidan} et~al.(2011){Pontoppidan}, {Blake}, and
  {Smette}}]{Pontoppidan11}
{Pontoppidan} KM, {Blake} GA {Smette} A (2011) {The Structure and Dynamics of
  Molecular Gas in Planet-forming Zones: A CRIRES Spectro-astrometric Survey}.
  \apj 733:84

\bibitem[{{Preibisch} et~al.(2005){Preibisch}, {Kim}, {Favata}, {Feigelson},
  {Flaccomio}, {Getman}, {Micela}, {Sciortino}, {Stassun}, {Stelzer}, and
  {Zinnecker}}]{Preibisch05}
{Preibisch} T, {Kim} YC, {Favata} F et~al. (2005) {The Origin of T Tauri X-Ray
  Emission: New Insights from the Chandra Orion Ultradeep Project}. \apjs
  160:401--422

\bibitem[{{Qi} et~al.(2013){Qi}, {Oberg}, {Wilner}, {d'Alessio}, {Bergin},
  {Andrews}, {Blake}, {Hogerheijde}, and {van Dishoeck}}]{Qi13_sci}
{Qi} C, {Oberg} KI, {Wilner} DJ et~al. (2013) {Imaging of the CO Snow Line in a
  Solar Nebula Analog}. Science 341:630--

\bibitem[{{Qi} et~al.(2015){Qi}, {{\"O}berg}, {Andrews}, {Wilner}, {Bergin},
  {Hughes}, {Hogherheijde}, and {D'Alessio}}]{qi2015}
{Qi} C, {{\"O}berg} KI, {Andrews} SM et~al. (2015) {Chemical Imaging of the CO
  Snow Line in the HD 163296 Disk}. \apj 813:128

\bibitem[{{Reboussin} et~al.(2015){Reboussin}, {Wakelam}, {Guilloteau},
  {Hersant}, and {Dutrey}}]{Reboussin15}
{Reboussin} L, {Wakelam} V, {Guilloteau} S, {Hersant} F {Dutrey} A (2015)
  {Chemistry in protoplanetary disks: the gas-phase CO/H$_{2}$ ratio and the
  carbon reservoir}. \aap 579:A82

\bibitem[{{Salmeron} et~al.(2007){Salmeron}, {K{\"o}nigl}, and
  {Wardle}}]{salmeron2007}
{Salmeron} R, {K{\"o}nigl} A {Wardle} M (2007) {Angular momentum transport in
  protostellar discs}. \mnras 375:177--183

\bibitem[{{Salyk} et~al.(2008){Salyk}, {Pontoppidan}, {Blake}, {Lahuis}, {van
  Dishoeck}, and {Evans}}]{salyk08}
{Salyk} C, {Pontoppidan} KM, {Blake} GA et~al. (2008) {H$_{2}$O and OH Gas in
  the Terrestrial Planet-forming Zones of Protoplanetary Disks}. \apjl
  676:L49--L52

\bibitem[{{Salyk} et~al.(2011){Salyk}, {Pontoppidan}, {Blake}, {Najita}, and
  {Carr}}]{Salyk11}
{Salyk} C, {Pontoppidan} KM, {Blake} GA, {Najita} JR {Carr} JS (2011) {A
  Spitzer Survey of Mid-infrared Molecular Emission from Protoplanetary Disks.
  II. Correlations and Local Thermal Equilibrium Models}. \apj 731:130

\bibitem[{{Schindhelm} et~al.(2012){Schindhelm}, {France}, {Herczeg}, {Bergin},
  {Yang}, {Brown}, {Brown}, {Linsky}, and {Valenti}}]{Schindhelm12}
{Schindhelm} E, {France} K, {Herczeg} GJ et~al. (2012) {Ly{$\alpha$} Dominance
  of the Classical T Tauri Far-ultraviolet Radiation Field}. \apjl 756:L23

\bibitem[{{Schwarz} et~al.(2016){Schwarz}, {Bergin}, {Cleeves}, {Blake},
  {Zhang}, {{\"O}berg}, {van Dishoeck}, and {Qi}}]{Schwarz16}
{Schwarz} KR, {Bergin} EA, {Cleeves} LI et~al. (2016) {The Radial Distribution
  of H$_{2}$ and CO in TW Hya as Revealed by Resolved ALMA Observations of CO
  Isotopologues}. \apj 823:91

\bibitem[{{Shakura} and {Syunyaev}(1973)}]{ss73}
{Shakura} NI {Syunyaev} RA (1973) {Black holes in binary systems. Observational
  appearance.} \aap 24:337--355

\bibitem[{{Simon} et~al.(2015){Simon}, {Hughes}, {Flaherty}, {Bai}, and
  {Armitage}}]{simon2015}
{Simon} JB, {Hughes} AM, {Flaherty} KM, {Bai} XN {Armitage} PJ (2015)
  {Signatures of MRI-driven Turbulence in Protoplanetary Disks: Predictions for
  ALMA Observations}. \apj 808:180

\bibitem[{{Simon} et~al.(2017){Simon}, {Bai}, {Flaherty}, and
  {Hughes}}]{simon2017}
{Simon} JB, {Bai} XN, {Flaherty} KM {Hughes} AM (2017) {A New Model for Weak
  Turbulence in Protoplanetary Disks}. ArXiv e-prints

\bibitem[{{Sofia} et~al.(1994){Sofia}, {Cardelli}, and {Savage}}]{scs94}
{Sofia} UJ, {Cardelli} JA {Savage} BD (1994) {The abundant elements in
  interstellar dust}. \apj 430:650--666

\bibitem[{{Stammler} et~al.(2017){Stammler}, {Birnstiel}, {Pani{\'c}},
  {Dullemond}, and {Dominik}}]{stammler2017}
{Stammler} SM, {Birnstiel} T, {Pani{\'c}} O, {Dullemond} CP {Dominik} C (2017)
  {Redistribution of CO at the location of the CO ice line in evolving gas and
  dust disks}. \aap 600:A140

\bibitem[{{Stevenson} and {Lunine}(1988)}]{Stevenson88}
{Stevenson} DJ {Lunine} JI (1988) {Rapid formation of Jupiter by diffuse
  redistribution of water vapor in the solar nebula}. \icarus 75:146--155

\bibitem[{{Suzuki} and {Inutsuka}(2009)}]{suzuki2009}
{Suzuki} TK {Inutsuka} Si (2009) {Disk Winds Driven by Magnetorotational
  Instability and Dispersal of Protoplanetary Disks}. \apjl 691:L49--L54

\bibitem[{{Teague} et~al.(2016){Teague}, {Guilloteau}, {Semenov}, {Henning},
  {Dutrey}, {Pi{\'e}tu}, {Birnstiel}, {Chapillon}, {Hollenbach}, and
  {Gorti}}]{Teague16}
{Teague} R, {Guilloteau} S, {Semenov} D et~al. (2016) {Measuring turbulence in
  TW Hydrae with ALMA: methods and limitations}. \aap 592:A49

\bibitem[{{Tennyson}(2011)}]{Tennyson11}
{Tennyson} J (2011) {Astronomical Spectroscopy: AN Introduction to the Atomic
  and Molecular Physics of Astronomical Spectra (2ND Edition)}. World
  Scientific, \doi{10.1142/7574}

\bibitem[{{Thi} et~al.(2010){Thi}, {Mathews}, {M{\'e}nard}, {Woitke}, {Meeus},
  {Riviere-Marichalar}, {Pinte}, {Howard}, {Roberge}, {Sandell}, {Pascucci},
  {Riaz}, {Grady}, {Dent}, {Kamp}, {Duch{\^e}ne}, {Augereau}, {Pantin},
  {Vandenbussche}, {Tilling}, {Williams}, {Eiroa}, {Barrado}, {Alacid},
  {Andrews}, {Ardila}, {Aresu}, {Brittain}, {Ciardi}, {Danchi}, {Fedele}, {de
  Gregorio-Monsalvo}, {Heras}, {Huelamo}, {Krivov}, {Lebreton}, {Liseau},
  {Martin-Zaidi}, {Mendigut{\'{\i}}a}, {Montesinos}, {Mora},
  {Morales-Calderon}, {Nomura}, {Phillips}, {Podio}, {Poelman}, {Ramsay},
  {Rice}, {Solano}, {Walker}, {White}, and {Wright}}]{thi10}
{Thi} WF, {Mathews} G, {M{\'e}nard} F et~al. (2010) {Herschel-PACS observation
  of the 10 Myr old T Tauri disk TW Hya. Constraining the disk gas mass}. \aap
  518:L125

\bibitem[{{Thiemens}(2006)}]{Thiemens06}
{Thiemens} MH (2006) {History and Applications of Mass-Independent Isotope
  Effectsa}. Annual Review of Earth and Planetary Sciences 34:217--262

\bibitem[{{Tsukagoshi} et~al.(2015){Tsukagoshi}, {Momose}, {Saito}, {Kitamura},
  {Shimajiri}, and {Kawabe}}]{Tsukagoshi15}
{Tsukagoshi} T, {Momose} M, {Saito} M et~al. (2015) {First Detection of [C I]
  $^{3}$P$_{1}$-$^{3}$P$_{0}$ Emission from a Protoplanetary Disk}. \apjl
  802:L7

\bibitem[{{Umebayashi} and {Nakano}(1981)}]{Umebayashi81}
{Umebayashi} T {Nakano} T (1981) {Fluxes of Energetic Particles and the
  Ionization Rate in Very Dense Interstellar Clouds}. \pasj 33:617

\bibitem[{{van Dishoeck} et~al.(2013{\natexlab{a}}){van Dishoeck}, {Herbst},
  and {Neufeld}}]{vandishoeckH2Oreview}
{van Dishoeck} EF, {Herbst} E {Neufeld} DA (2013{\natexlab{a}}) {Interstellar
  Water Chemistry: From Laboratory to Observations}. Chemical Reviews
  113:9043--9085

\bibitem[{{van Dishoeck} et~al.(2013{\natexlab{b}}){van Dishoeck}, {Herbst},
  and {Neufeld}}]{vanDishoeck13}
{van Dishoeck} EF, {Herbst} E {Neufeld} DA (2013{\natexlab{b}}) {Interstellar
  Water Chemistry: From Laboratory to Observations}. Chemical Reviews
  113:9043--9085

\bibitem[{{van Zadelhoff} et~al.(2001){van Zadelhoff}, {van Dishoeck}, {Thi},
  and {Blake}}]{vanz_etal01}
{van Zadelhoff} GJ, {van Dishoeck} EF, {Thi} WF {Blake} GA (2001)
  {Submillimeter lines from circumstellar disks around pre-main sequence
  stars}. \aap 377:566--580

\bibitem[{{van't Hoff} et~al.(2017){van't Hoff}, {Walsh}, {Kama}, {Facchini},
  and {van Dishoeck}}]{vanthoff2017}
{van't Hoff} MLR, {Walsh} C, {Kama} M, {Facchini} S {van Dishoeck} EF (2017)
  {Robustness of N$_{2}$H$^{+}$ as tracer of the CO snowline}. \aap 599:A101

\bibitem[{{Visser} et~al.(2009){Visser}, {van Dishoeck}, and
  {Black}}]{visser09}
{Visser} R, {van Dishoeck} EF {Black} JH (2009) {The photodissociation and
  chemistry of CO isotopologues: applications to interstellar clouds and
  circumstellar disks}. \aap 503:323--343

\bibitem[{{Wagner} and {Graff}(1987)}]{wg87}
{Wagner} AF {Graff} MM (1987) {Oxygen chemistry of shocked interstellar clouds.
  I - Rate constants for thermal and nonthermal internal energy distributions}.
  \apj 317:423--431

\bibitem[{{Wang} et~al.(2005){Wang}, {Bell}, {Iedema}, {Tsekouras}, and
  {Cowin}}]{wang2005}
{Wang} H, {Bell} RC, {Iedema} MJ, {Tsekouras} AA {Cowin} JP (2005) {Sticky Ice
  Grains Aid Planet Formation: Unusual Properties of Cryogenic Water Ice}. \apj
  620:1027--1032

\bibitem[{{Webber}(1998)}]{webber_cosmicray}
{Webber} WR (1998) {A New Estimate of the Local Interstellar Energy Density and
  Ionization Rate of Galactic Cosmic Cosmic Rays}. \apj 506:329--334

\bibitem[{{Weidenschilling}(1977)}]{w77_mmsn}
{Weidenschilling} SJ (1977) {The distribution of mass in the planetary system
  and solar nebula}. \apss 51:153--158

\bibitem[{{Weidenschilling} and {Cuzzi}(1993)}]{wc_ppiii}
{Weidenschilling} SJ {Cuzzi} JN (1993) {Formation of planetesimals in the solar
  nebula}. In: {Levy} EH {Lunine} JI (eds) Protostars and Planets III, pp
  1031--1060

\bibitem[{{Whipple}(1973)}]{Whipple73}
{Whipple} FL (1973) {Radial Pressure in the Solar Nebula as Affecting the
  Motions of Planetesimals}. NASA Special Publication 319:355

\bibitem[{{Woitke} et~al.(2009){Woitke}, {Kamp}, and {Thi}}]{woitke09}
{Woitke} P, {Kamp} I {Thi} WF (2009) {Radiation thermo-chemical models of
  protoplanetary disks. I. Hydrostatic disk structure and inner rim}. \aap
  501:383--406

\bibitem[{{Wong} et~al.(2008){Wong}, {Lunine}, {Atreya}, {Johnson}, {Mahaffy},
  {Owen}, and {Encrenaz}}]{wong2008}
{Wong} MH, {Lunine} JI, {Atreya} S et~al. (2008) {conference two}. in Reviews
  in Mineralogy and Geochemistry: Oxygen in the Earliest Solar System, (eds G
  McPherson et al), Mineralogical Society of America, 68, 219

\bibitem[{{Yang} et~al.(2012){Yang}, {Herczeg}, {Linsky}, {Brown},
  {Johns-Krull}, {Ingleby}, {Calvet}, {Bergin}, and {Valenti}}]{Yang12}
{Yang} H, {Herczeg} GJ, {Linsky} JL et~al. (2012) {A Far-ultraviolet Atlas of
  Low-resolution Hubble Space Telescope Spectra of T Tauri Stars}. \apj 744:121

\bibitem[{{Yoshino} et~al.(1996){Yoshino}, {Esmond}, {Parkinson}, {Ito}, and
  {Matsui}}]{yoshino_h2oxs}
{Yoshino} K, {Esmond} JR, {Parkinson} WH, {Ito} K {Matsui} T (1996) {Absorption
  Cross Section Measurements of Water Vapor in the Wavelength Region 120 nm to
  188 nm.} Chemical Physics 211:387--+

\bibitem[{{Zhang} et~al.(2013){Zhang}, {Pontoppidan}, {Salyk}, and
  {Blake}}]{zhang13}
{Zhang} K, {Pontoppidan} KM, {Salyk} C {Blake} GA (2013) {Evidence for a Snow
  Line beyond the Transitional Radius in the TW Hya Protoplanetary Disk}. \apj
  766:82

\bibitem[{{Zhang} et~al.(2017){Zhang}, {Bergin}, {Blake}, {Cleeves}, and
  {Schwarz}}]{Zhang17}
{Zhang} K, {Bergin} EA, {Blake} GA, {Cleeves} L {Schwarz} K (2017) {Unvieling
  the mass inventory of the giant-planet formation zone in a solar nebula
  analog}. Nature Astronomy p submitted

\end{thebibliography}

\end{document}